\documentclass[twocolumn,aps,prb, 
superscriptaddress, longbibliography]{revtex4-2}
\usepackage{amsmath,latexsym}
\usepackage{xcolor}
\usepackage{natbib}
\usepackage[%
  colorlinks=true,
  urlcolor=blue,
  linkcolor=blue,
  citecolor=blue
]{hyperref}
\usepackage{etoolbox}
\usepackage{breqn}
\makeatletter
\let\cat@comma@active\@empty
\makeatletter
\usepackage{graphicx}
\usepackage{dcolumn}
\usepackage{bm}%
\usepackage{dcolumn}
\usepackage{siunitx}
\usepackage{amssymb}
\usepackage{booktabs}
\usepackage{setspace}
\usepackage{xcolor}
\usepackage{amsmath}
\usepackage{amsthm}
\usepackage{natbib}
\usepackage{appendix}
\usepackage{textgreek}
\usepackage{color}
\usepackage{orcidlink}

\begin{document}

\title{{Phonon thermal Hall effect in non-magnetic {Y${\rm _2}$Ti${\rm _2}$O${\rm _7}$}}}

\author{Rohit Sharma\,\orcidlink{0000-0001-9815-0733}}
\email{sharma@ph2.uni-koeln.de}
\affiliation{II.\, Physikalisches Institut, Universit\"at zu K\"oln, Z\"ulpicher Str.\ 77, 50937 K\"oln, Germany}
\author{Martin Valldor\,\orcidlink{0000-0001-7061-3492}}
\affiliation{II.\, Physikalisches Institut, Universit\"at zu K\"oln, Z\"ulpicher Str.\ 77, 50937 K\"oln, Germany}
\affiliation{Centre for Materials Science and Nanotechnology (SMN), Department of Chemistry,
University of Oslo, Oslo N-0371, Norway}
\author{Thomas Lorenz\,\orcidlink{0000-0003-4832-5157}}
\email{tl@ph2.uni-koeln.de}
\affiliation{II.\, Physikalisches Institut, Universit\"at zu K\"oln, Z\"ulpicher Str.\ 77, 50937 K\"oln, Germany}
\date{\today}

\begin{abstract}
 We report an investigation of the phonon thermal Hall effect in single crystal samples of Y$_2$Ti$_2$O$_7$, Dy$_2$Ti$_2$O$_7$, and DyYTi$_2$O$_7$. We measured the field-linear thermal Hall effect in all three samples. The temperature dependence of thermal Hall conductivities shows a peak around 15~K, which coincides with the peak positions of the longitudinal thermal conductivities. The temperature-dependent longitudinal thermal conductivities indicates that phonons dominate thermal transport in all three samples. However, the presence of Dy$^{3+}$ magnetic ions introduces significant effects on the field dependence of the longitudinal thermal conductivities. The thermal Hall ratio is sizeable in all three samples and consistent with the values reported for other insulating materials exhibiting a phononic thermal Hall effect, though their exact underlying mechanism remains yet to be identified. The thermal Hall ratio is nearly the same for Y$_2$Ti$_2$O$_7$ and DyYTi$_2$O$_7$, and slightly larger for Dy$_2$Ti$_2$O$_7$, suggesting that magnetic impurities are less significant in generating the phononic thermal Hall effect. Our observations of the phononic thermal Hall effect support an intrinsic origin in Y$_2$Ti$_2$O$_7$, and suggest a combination of intrinsic and extrinsic effects in Dy$_2$Ti$_2$O$_7$ and DyYTi$_2$O$_7$.
 
 \end{abstract}

\date{\today}
\maketitle

The thermal Hall effect (THE) of phonons observed in various insulating materials continues to be a subject of ongoing debate, particularly regarding its origin and the intrinsic versus extrinsic nature of its underlying mechanisms. Intrinsic mechanisms involve phonon dispersion, including theoretical scenarios such as Berry curvature in phonon bands \cite{2012PhRvB..86j4305Q, 2019PhRvL.123y5901S, 2012PhRvB..85m4411I}, phonon scattering due to collective fluctuations \cite{2022PhRvB.106x5139M, 2022PhRvX..12d1031M}, and interactions between phonons and other quasi-particles like magnons \cite{2021PhRvB.104c5103Z, 2019PhRvL.123p7202Z, 2006PhRvL..96o5901S, 2016PhRvL.117u7205T}. On the other hand, extrinsic mechanisms invoke scattering of
phonons by charged impurities or defects \cite{{2022PNAS..11915141G},{2021PhRvB.103t5115G},{2022PhRvB.105v0301F},{2020PhRvL.124p7601C},{2022PhRvB.106n4111S},{2014PhRvL.113z5901M, {PhysRevResearch.5.033197}}}. Experimentally, phononic THE was first detected in the paramagnetic insulator Tb$_3$Ga$_5$O$_{12}$ \cite{2005PhRvL..95o5901S}, followed by observations in various other insulating solids \cite{2019Natur.571..376G, 2020NatPh..16.1108G, 2022PNAS..11908016C, 2020NatCo..11.5325B, 2017PhRvL.118n5902S, 2017NatMa..16..797I, 2022NatCo..13.4604U, 
2015Sci...348..106H, 2017PhRvL.118n5902S, 2019PhRvB..99m4419H, 2021PhRvL.126a5901S, 2015PhRvL.115j6603H, 2024NatCo..15..243K,  
2019PhRvB..99h5136H, 2021PhRvL.127x7202Z, 2022PhRvB.105k5101B, 2023PhRvB.107f0404X, 2022PhRvX..12b1025L, 2022PNAS..11901975J, 2024arXiv240313306M, 2023PhRvR...5d3110G, 2023PhRvB.108n0402L, 2010Sci...329..297O, Ataei_2024, 2023PhRvB.108n0402L, 2020PhRvX..10d1059A}, including non-magnetic materials \cite{2020PhRvL.124j5901L, 2023NatCo..14.1027L, 2024PhRvB.109j4304S, 2024arXiv240402863J, 2024arXiv240402863J}. The recent observation of phonon THE in several non-magnetic elemental solids, such as the insulator black phosphorus \cite{2023NatCo..14.1027L}, and semiconductors like Si and Ge \cite{2024arXiv240402863J}, has pointed toward the universality of phonon THE in solids. Additionally, the thermal Hall ratio has been phenomenologically shown to exhibit a universal scaling behavior observed across various classes of insulating materials, including non-magnetic ones \cite{2023NatCo..14.1027L}.

The observation of phonon THE across a wide variety of insulating solids prompted us to re-examine THE in insulating pyrochlore Y$_2$Ti$_2$O$_7$, where a previous report by Hirschberger et al. \cite{2015Sci...348..106H} indicated a negligible thermal Hall ratio. Another motivation for this revisit is that the study by Hirschberger et al. was focused more on the isostructural magnetic material Tb$_2$Ti$_2$O$_7$, where a large THE was measured and attributed to exotic neutral excitations associated with its quantum spin liquid state. The thermal Hall ratio measured in Y$_2$Ti$_2$O$_7$ was only reported at 15~K, and lead to the conclusion that Y$_2$Ti$_2$O$_7$ exhibits null signal, in comparison to the significant THE observed in Tb$_2$Ti$_2$O$_7$. Later, Hirokane et al. \cite{2019PhRvB..99m4419H} interpreted the observed large THE in Tb$_2$Ti$_2$O$_7$ to be of  phononic origin, because they measured a comparable thermal Hall signal in a diluted sample in which 70\% of the Tb$^{3+}$ ions were replaced by non-magnetic Y$^{3+}$ ions. This naturally raises the question whether the pure Y$_2$Ti$_2$O$_7$ in fact exhibits a null signal or a finite phononic THE, but up to now we are not aware of any systematic study of the THE in Y$_2$Ti$_2$O$_7$.

\begin{figure*}[tbp]
	\centering
	\hspace*{-0.6cm}
	\includegraphics[width=2.1\columnwidth]{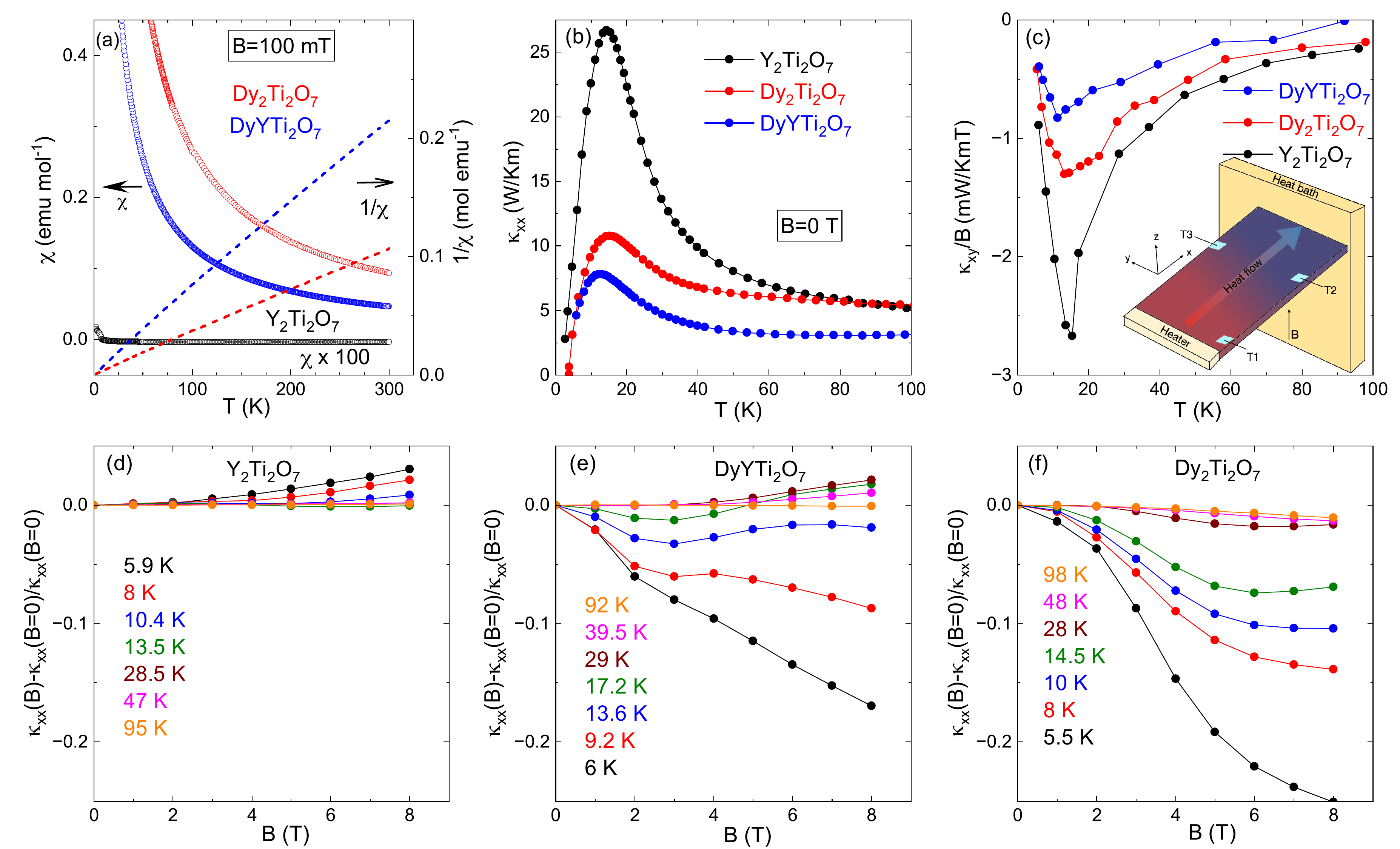} 
	\caption{
		(a) Magnetic susceptibility of Dy$_2$Ti$_2$O$_7$ and DyYTi$_2$O$_7$ showing Curie-Weiss behavior in comparison to Y$_2$Ti$_2$O$_7$, which is nonmagnetic apart from the temperature independent Core diamagnetism and a tiny impurity contribution (note that $\chi$ of Y$_2$Ti$_2$O$_7$ is   multiplied by 100). (b) Thermal conductivities $\kappa_{xx}$ of Y$_2$Ti$_2$O$_7$, Dy$_2$Ti$_2$O$_7$, and DyYTi$_2$O$_7$ measured in zero magnetic field and (c) the corresponding Thermal Hall conductivities normalized by the field, that is $\kappa_{xy}/B$. Panel (c) also displays the schematic setup to measure $\kappa_{xx}$ and $\kappa_{xy}$ from the longitudinal and transverse temperature differences ($\Delta T_x = T_1 - T_2$, $\Delta T_y = T_2 - T_3$) induced by a heat current $J_Q \parallel x$ and a magnetic field $B   \parallel z$. With respect to the cubic lattice directions, $J_Q$ was along $[110]$ and $B \parallel [001]$ in Y$_2$Ti$_2$O$_7$ and Dy$_2$Ti$_2$O$_7$, whereas in DyYTi$_2$O$_7$ the orientations $J_Q \parallel [001]$ and $B \parallel [110]$ were used. (d-f) Magnetic field dependences of $\kappa_{xx}$ for all three samples, which remains within the range of 3\% for Y$_2$Ti$_2$O$_7$, but reaches up to 25\% for Dy$_2$Ti$_2$O$_7$ at the lowest measured temperature. } 
	\label{FIG.1.}
\end{figure*}

In this letter, we report the discovery of a sizeable phonon THE in the non-magnetic insulator Y$_2$Ti$_2$O$_7$ from a comparative study of THE in Y$_2$Ti$_2$O$_7$ alongside the
isostructural pyrochlore materials Dy$_2$Ti$_2$O$_7$ and DyYTi$_2$O$_7$, which are paramagnetic insulators with large local moments of the Dy$^{3+}$ ions, see Fig.~\ref{FIG.1.}.
Our discovery of phonon THE in Y$_2$Ti$_2$O$_7$ and the related magnetic materials aligns well with observations of the phononic THE measured in various insulating solids \cite{2019Natur.571..376G, 2020NatPh..16.1108G, 2020NatCo..11.5325B, 2022PNAS..11908016C, 2021PhRvL.126a5901S, 2019PhRvB..99m4419H, 2022PhRvB.105k5101B, 2022PNAS..11901975J, 2022PhRvX..12b1025L, Ataei_2024, 2020PhRvL.124j5901L, 2023NatCo..14.1027L, 2024PhRvB.109j4304S, 2024arXiv240402863J}. It is noteworthy that Dy$_2$Ti$_2$O$_7$ is well-known for spin-ice physics, which involves magnetic monopole excitations resulting from magnetic frustration that prevents the formation of long-range magnetic order \cite{2008Natur.451...42C, 2009Sci...326..411M, 2014NatPh..10..135P, 2001Sci...294.1495B}. The magnetic monopole excitations and their impact on various physical properties (specific heat, thermal conductivity, and magnetization) \cite{pomaranski2013absence, ramirez1999zero, 2013PhRvB..88e4406K, 2012PhRvB..86f0402K, scharffe2015heat, 2015PhRvB..92r0405S, 2002PhRvB..65e4410F, PhysRevB.98.134446}, are relevant only at low temperatures ($T<10\,$K) and will not be in the focus of this study.

Single crystals of Y$_2$Ti$_2$O$_7$, Dy$_2$Ti$_2$O$_7$, and DyYTi$_2$O$_7$ were grown using the floating-zone technique, starting from sintered bars of TiO$_2$, Y$_2$O$_3$, and Dy$_2$O$_3$ mixed in the appropriate stoichiometries. The cool-pressed mixtures were directly introduced in the floating-zone furnace and were rapidly run through the melting zone to achieve a prime reaction. Subsequently, the feeding rod and the seed switched places before a slow crystal growth process began, at speeds of typically about $1\,$mm/h, see also Refs.~\cite{2013PhRvB..88e4406K, 2015PhRvB..92r0405S}. The thermal conductivities were measured using the standard steady-state, 1-heater 3-thermometer technique under high-vacuum conditions. We employed the steady-state technique where the magnetic field was changed in steps of $1\,$T from  $-8\,$T to  $8\,$T while maintaining the sample temperature at a constant set point. We waited at least 300 seconds after each incremental change of the magnetic field before taking measurements to avoid  transient contaminations caused by magneto-caloric and eddy-current heating effects. The temperature differences, $\Delta T_x = T_1 - T_2$ and $\Delta T_y = T_2 - T_3$, were produced by a  $10\,$k$\Omega$ RuO$_2$ chip resistor (heater) attached at one end of the sample and were measured with magnetic- field calibrated Cernox sensors (CX-1070), as sketched in Fig.~\ref{FIG.1.}(c). In the case of Y$_2$Ti$_2$O$_7$ and Dy$_2$Ti$_2$O$_7$, the heat current $J_Q = RI^2$ was applied along the [110] direction in a perpendicular magnetic field $B\parallel [001]$. For DyYTi$_2$O$_7$, the heat current and magnetic field directions were interchanged, i.e., $J_Q \parallel [001]$ and $B \parallel [110]$. Gold wires were used for making contact on the sample with silver paste and were connected to their respective thermometers of a home-built setup for thermal transport measurements. To eliminate the  misalignment of the transverse contacts, the temperature difference $\Delta T_H$ was obtained by antisymmetrization of the respective raw data measured in $\pm B$, i.e., $\Delta T_H (B) = (\Delta T_y (T, H) - \Delta T_y (T, -H)) / 2$. The thermal Hall conductivity is then obtained as $\kappa_{xy} = (\Delta T_H / \Delta T_x)  (l / w)  \kappa_{xx}$ with longitudinal thermal conductivity $\kappa_{xx} = (J_Q / \Delta T_x)  (l / wt)$, distance $l$ between the longitudinal contacts, sample width $w$, and thickness $t$.

\begin{figure*}[!tbp]
	\centering
	\hspace*{-0.6cm}
    \includegraphics[width=2.0\columnwidth]{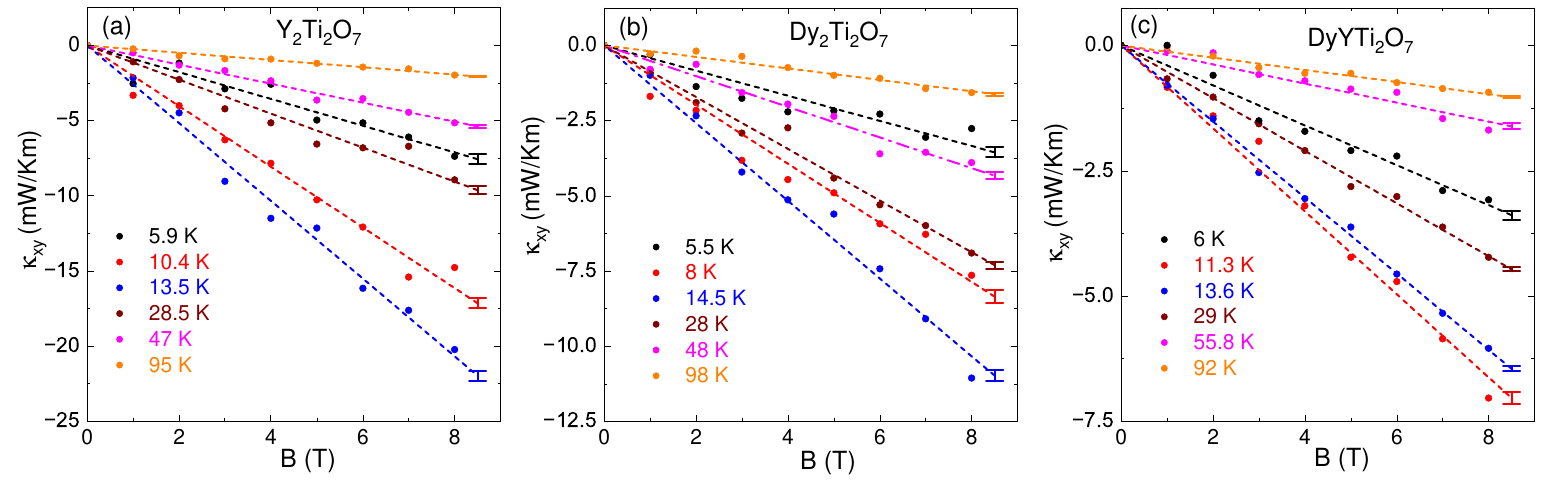}
	\caption{(a-c) Field-antisymmetrized thermal Hall conductivities $\kappa_{xy}$(B) measured at constant temperatures on Y$_2$Ti$_2$O$_7$, Dy$_2$Ti$_2$O$_7$, and DyYTi$_2$O$_7$. The dashed lines are linear fits that were used to calculate the temperature-dependent $\kappa_{xy}/B$. 
	For each fit, an error bar is calculated for $B=8.5\,$T using the standard deviation error of the obtained slope.}   
	\label{FIG.2.}
\end{figure*}
 
The temperature dependence of the longitudinal thermal conductivity ($\kappa_{xx}$), as shown in Fig.~\ref{FIG.1.}(b), measured on Y$_2$Ti$_2$O$_7$, Dy$_2$Ti$_2$O$_7$, and DyYTi$_2$O$_7$, exhibits typical phonon behavior with a peak around 15~K. The increase in $\kappa_{xx}$ at low temperatures, followed by its decrease above the peak temperature, results from the opposite temperature dependencies of phonon heat capacity ($C$) and phonon mean-free path ($l$) \cite{berman1976thermal}. Based on kinetic theory, the thermal conductivity \( \kappa = \frac{1}{3} C v \ell \), where \( C \), \( v \), and \( \ell \) represent the phonon specific heat, sound velocity, and mean free path, respectively. In insulating solids, the phonon heat conductivity is expected to vary as \( T^3 \) in the low-temperature limit, whereas \( \frac{1}{T} \) is approached at high temperatures. This arises from the temperature dependencies of \( C \) and \( \ell \), which approach \( T^3 \) or \( \frac{1}{T} \) behavior in the respective temperature regimes \cite{kittel2018introduction}, and \( \kappa_{xx}(T) \) is governed by the dominating temperature dependence of either \( C \) or \( \ell \). The absolute values of $\kappa_{xx}$ for Y$_2$Ti$_2$O$_7$ and Dy$_2$Ti$_2$O$_7$ are the same at higher temperatures, whereas the peak value for Y$_2$Ti$_2$O$_7$ ($\approx 27\,$W/Km) is significantly larger than for Dy$_2$Ti$_2$O$_7$ ($\approx 12\,$W/Km). The measured $\kappa_{xx}(T)$ for both, Y$_2$Ti$_2$O$_7$ and Dy$_2$Ti$_2$O$_7$, are in good agreement with those previously reported~\cite{2012PhRvB..86f0402K, 2019Natur.571..376G, 2013PhRvB..87u4408L, 2015PhRvB..92i4408L}. The  $\kappa_{xx}(T)$ for DyYTi$_2$O$_7$ remains smaller than the other two across the entire temperature range and shows a peak value of $\approx 8\,$W/Km. The reduced $\kappa_{xx}(T)$ values in DyYTi$_2$O$_7$ compared to the pure compounds are likely due to the additional disorder introduced by the statistical distribution of Dy and Y ions. The fact that Y$_2$Ti$_2$O$_7$ has a larger thermal conductivity peak than Dy$_2$Ti$_2$O$_7$ can be attributed to additional presence of magnetic scattering in the latter, which leads to extra suppression of $\kappa_{xx}$. Y$_2$Ti$_2$O$_7$ is non-magnetic, whereas Dy$_2$Ti$_2$O$_7$ has a local magnetic moment  $\mu \simeq 10\,\mu_{\rm B}$/Dy$^{3+}$, in agreement with Hund's rules. Therefore, phonon scattering due to magnetic Dy$^{3+}$ ions could also be a potential source of suppressed $\kappa_{xx}$ at low temperatures for Dy$_2$Ti$_2$O$_7$ and DyYTi$_2$O$_7$ samples. The strong suppression of $\kappa_{xx}$ for DyYTi$_2$O$_7$ is the combined effect of disorder induced by the random distribution of Dy and Y ions, and the magnetic scattering of phonons compared to the parent compound Dy$_2$Ti$_2$O$_7$.

\begin{figure*}[t]
	\centering
 \includegraphics[width=2.0\columnwidth]{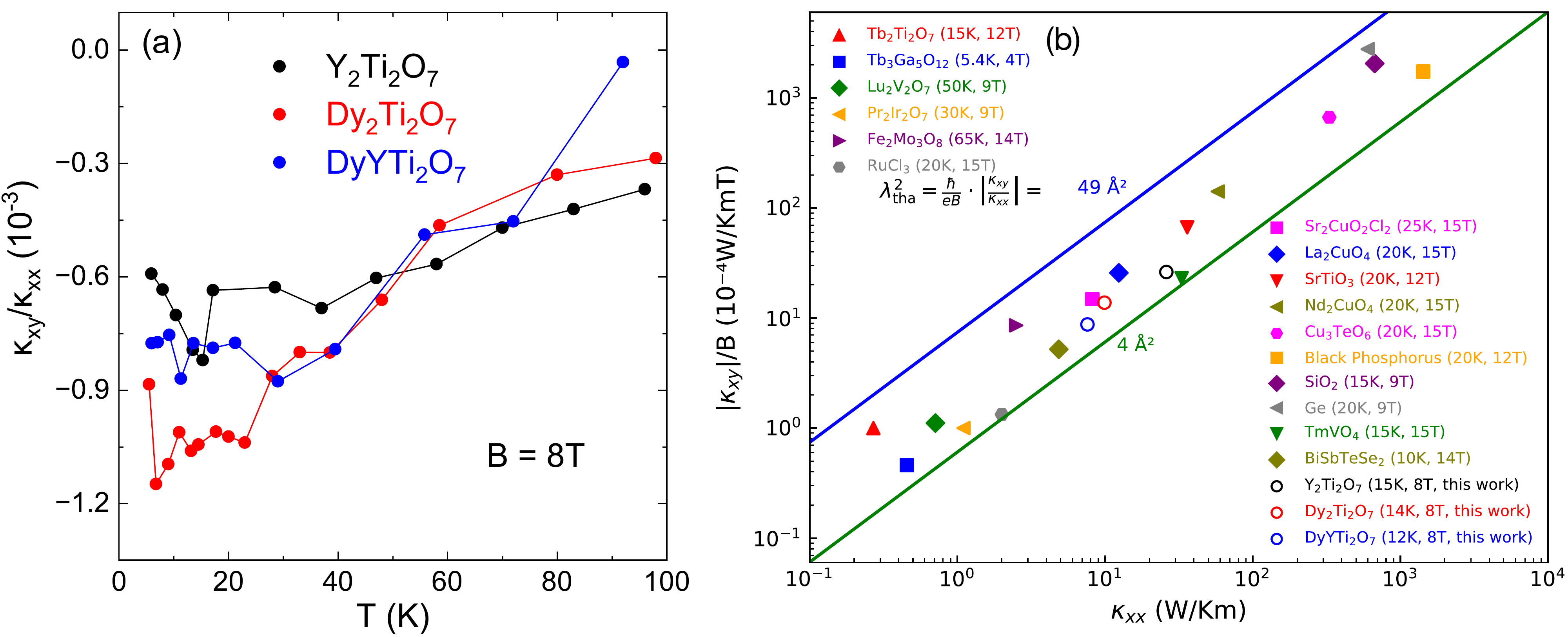} 
	\caption{ (a) Temperature-dependent thermal Hall ratios $\kappa_{xy}/\kappa_{xx}$ for Y$_2$Ti$_2$O$_7$, Dy$_2$Ti$_2$O$_7$, and DyYTi$_2$O$_7$. The maximum value of $\kappa_{xy}/\kappa_{xx}$ remains in the range of $-10^{-4}$ to $-10^{-3}$. (b) Scaling behaviour of thermal Hall conductivities normalized by the field, $|\kappa_{xy}|/B$, and longitudinal thermal conductivities $\kappa_{xx}$ of various insulators \cite{2005PhRvL..95o5901S, 2019PhRvB..99m4419H, 2017NatMa..16..797I, 2020NatPh..16.1108G, 2020NatCo..11.5325B, 2022PhRvX..12b1025L, 2020PhRvL.124j5901L, 2023NatCo..14.1027L, 2022NatCo..13.4604U, 2022PNAS..11908016C, 2024PhRvB.109j4304S, 2024arXiv240402863J, 2023arXiv231010643V5}. Using the magnetic length \( \ell_B = \sqrt{\frac{\hbar}{eB}} \), the extracted length scale \( \lambda_{\text{tha}} \) from the relation \( \lambda_{\text{tha}}^2 / \ell_B^2 = \kappa_{xy} / \kappa_{xx} \) remains between 2\AA{} to 7\AA{} for various insulators, despite their mean free paths varying by orders of magnitude.}
	\label{FIG.3.}
\end{figure*}

Figure~\ref{FIG.1.}(d-f) shows $\kappa_{xx}$ as a function of the magnetic field at different temperatures. For Y$_2$Ti$_2$O$_7$, the field dependence is negligible at higher temperatures and very weak (below about 3\%) at low temperatures. In contrast, Dy$_2$Ti$_2$O$_7$ exhibits the strongest relative field dependence, reaching about 25\% at the lowest measured temperature. DyYTi$_2$O$_7$ shows a relative field dependence of around 17\%. This strong field dependence in Dy$_2$Ti$_2$O$_7$ suggests that Dy$^{3+}$ ions play a significant role in the suppression of $\kappa_{xx}$, consistent with previous findings \cite{2015PhRvB..92i4408L, 2013PhRvB..88e4406K, 2004PhRvL..93q7202S}. Similar behavior and variation in the field dependence of $\kappa_{xx}$ have been observed in the paramagnetic insulator TmVO$_4$ when the magnetic Tm$^{3+}$ ion is replaced by Y$^{3+}$ ions \cite{2023arXiv231010643V5}. Furthermore, if we examine the field-dependent $\kappa_{xx}$ of DyYTi$_2$O$_7$ at higher field values, above 15~K, it shows an increase in $\kappa_{xx}$ compared to its zero-field value. In magnetically ordered materials, magnetic fluctuations around the ordering temperature can cause strong scattering of phonons, significantly suppressing the phononic heat conduction \cite{2004PhRvL..93q7202S, 2007PhRvB..76i4418B}. The application of a magnetic field partially suppresses these magnetic fluctuations, leading to the enhancement of phonon thermal conductivity \cite{2016NatCo...710807T}. Dy$_2$Ti$_2$O$_7$ and DyYTi$_2$O$_7$ do, however, not exhibit an increase in thermal conductivity with increasing field. Instead, their thermal conductivity decreases significantly at low temperatures and moderate fields, followed by a saturation at higher fields. With increasing temperature the overall field dependence rapidly decreases and finally results in a slight increase of $\kappa_{xx}$ at higher fields and temperatures. This behavior can be related either to field-enhanced magnetic scattering on phonons \cite{2015PhRvB..92i4408L, 2013PhRvB..88e4406K, 2004PhRvL..93q7202S} or to magnetoelastic coupling, and/or the presence of low-energy optical modes that intersect both longitudinal and transverse acoustic phonon modes \cite{2016PhRvB..93u4308R}. These low-energy optical modes have been identified in various rare-earth pyrochlore materials, where their interference with acoustic phonon modes effectively reduces the overall lattice thermal conductivity \cite{2015AcMat..91..304L}. In contrast, Y$_2$Ti$_2$O$_7$ shows a slight low-temperature increase in $\kappa_{xx}(B)$ of about 3\%. 
At first glance, one might imagine this increase of $\kappa_{xx}(B)$ to be a consequence of possible magnetic impurities in Y$_2$Ti$_2$O$_7$. From the oxidation states Y$^{3+}_2$Ti$^{4+}_2$O$^{2-}_7$, magnetic impurities can arise when diamagnetic Ti$^{4+}$ with $3d^0$ configuration, changes its oxidation state to Ti$^{3+}$ with $3d^1$ and one Bohr magneton $\mu_{\rm B}$ \cite{2013PhRvB..87r4427R}. In order to investigate this possibility, we measured the magnetization of the single crystal used for the thermal transport measurements. However, on this small sample (6.3 mg), it was not possible to detect a reliable magnetization signal in our SQUID magnetometer. Thus, we measured the magnetization on a much larger piece (516 mg) from the same single crystal. As shown in Fig.~\ref{FIG.1.}, the magnetic susceptibility is almost constant and weakly diamagnetic ($\simeq -4\,\cdot 10^{-5}\,$emu/mol) above $100\,$K, then it slowly increases upon further cooling and reaches $\simeq 2\cdot 10^{-4}\,$emu/mol at $2\,$K. From a low-temperature magnetization curve up to $7\,$T (see~\cite{sharma_2024_12755212}), we estimate the concentration of magnetic impurities to be below 0.05\%. 
Therefore, we conclude that the slight low-temperature increase of $\kappa_{xx}$ in Y$_2$Ti$_2$O$_7$ cannot be attributed to magnetic impurities and is likely of intrinsic origin. Furthermore, we conclude that the scattering of phonons by Dy$^{3+}$ magnetic ions in Dy$_2$Ti$_2$O$_7$ and by both magnetic impurities and disorder in DyYTi$_2$O$_7$ is responsible for the significant suppression of $\kappa_{xx}$ compared to nonmagnetic Y$_2$Ti$_2$O$_7$ in zero field.  

We now shift our focus to the thermal Hall conductivity ($\kappa_{xy}$). Figure~\ref{FIG.2.}(a-c) shows linear magnetic field dependences of $\kappa_{xy}$ with negative signs for all samples and temperatures. From linear fits to the field-dependent $\kappa_{xy}(B)$ data measured at different temperatures, we derive the 
temperature-dependent $\kappa_{xy}/B$ that is shown in Figure~\ref{FIG.1.}(c). Note the negative sign of $\kappa_{xy}$ for all three samples across the entire temperature range, which is in contrast to the positive $\kappa_{xy}$ observed in Tb$_2$Ti$_2$O$_7$~\cite{2015Sci...348..106H, 2019PhRvB..99m4419H}. For each sample, the temperature dependent $\kappa_{xy}/B$ shows a peak with a maximum value around the same temperature where a peak in $\kappa_{xx}$ occurs.  This coincidence of peaks in $\kappa_{xx}$ and $\kappa_{xy}$ has been reported in several insulating materials \cite{2019Natur.571..376G, 2020NatPh..16.1108G, 2020NatCo..11.5325B, 2022PNAS..11908016C, 2021PhRvL.126a5901S, 2019PhRvB..99m4419H, 2022PhRvB.105k5101B, 2022PNAS..11901975J, 2022PhRvX..12b1025L,  Ataei_2024, 2020PhRvL.124j5901L, 2023NatCo..14.1027L, 2024PhRvB.109j4304S, 2024arXiv240402863J, 2024arXiv240402863J}. It is postulated that since $\kappa_{xx}$, being phonon-dominated, and given the concurrence of peak positions between $\kappa_{xx}$ and $\kappa_{xy}$, $\kappa_{xy}$ is also of phononic origin. To further check the phononic origin, we can examine the thermal Hall ratio ($\kappa_{xy}/\kappa_{xx}$), which represents the degree of handedness (chirality). It has been found that $|\kappa_{xy}|/\kappa_{xx}$ for numerous materials lies in the range of $10^{-4}$ to $10^{-3}$, even though the absolute values of $\kappa_{xx}$ and $\kappa_{xy}$ vary by orders of magnitude at their peak temperatures \cite{2019Natur.571..376G, 2020NatPh..16.1108G, 2020NatCo..11.5325B, 2022PNAS..11908016C, 2021PhRvL.126a5901S, 2019PhRvB..99m4419H, 2022PhRvB.105k5101B, 2022PNAS..11901975J, 2022PhRvX..12b1025L, Ataei_2024, 2020PhRvL.124j5901L, 2023NatCo..14.1027L, 2024PhRvB.109j4304S, 2024arXiv240402863J}. The temperature dependence of $\kappa_{xy}/\kappa_{xx}$ (see Figure~\ref{FIG.3.}(a)) for all three samples decreases with increasing temperature, showing a maximum value within the same range of $-10^{-4}$ to $-10^{-3}$ for a magnetic field of 8~T. It is worth noting that the $\kappa_{xy}/\kappa_{xx}$ measured for Y$_2$Ti$_2$O$_7$ is a sizable signal, contrary to what has been originally reported by Hirschberger et al. \cite{2015Sci...348..106H} and has been later cited as being zero by others \cite{2020NatCo..11.5325B, 2022PNAS..11908016C, 2023arXiv231010643V5}. The measured value of $\kappa_{xy}/\kappa_{xx}$ suggests that the THE observed in all three samples is of phononic origin. This brings us to the central question: What is the underlying mechanism that causes phononic THE in these systems? For most of these materials, the microscopic origin of the THE has been attributed to phonons through two general coupling mechanisms. These mechanisms are either intrinsic, where phonons couple to a field-sensitive mechanism inherent to the host material, or extrinsic, where phonons couple through impurities and defects \cite{martelli2024phonons}. It is interesting to note that the value of  $\kappa_{xy}/\kappa_{xx}$ is nearly the same for all three samples at temperatures above 50~K. 
Below 50~K, the ratio $\kappa_{xy}/\kappa_{xx}$ is largest in Dy$_2$Ti$_2$O$_7$ and smallest in Y$_2$Ti$_2$O$_7$, with the ratio in DyYTi$_2$O$_7$ only slightly exceeding that in Y$_2$Ti$_2$O$_7$. This rules out the possibility of a THE that is mainly arising from magnetic impurities. The enhanced $\kappa_{xy}/\kappa_{xx}$ in Dy$_2$Ti$_2$O$_7$ below 50~K indicates that skew-scattering on magnetic Dy$^{3+}$ ions has an additional impact, but this effect appears to be rather weak. Notably, a non-zero phononic THE has been observed in other non-magnetic insulators \cite{2020PhRvL.124j5901L, 2023NatCo..14.1027L, 2024arXiv240402863J, 2024arXiv240402863J}, and it has been argued that the phonon THE is an intrinsic property of solids, which is supported by our observation of a phononic THE in nonmagnetic Y$_2$Ti$_2$O$_7$. Thus, we further explore the applicability of this intrinsic scenario across all three samples.    

Recently, Li et al. \cite{2023NatCo..14.1027L} proposed a link between the universality of $|\kappa_{xy}|/\kappa_{xx}$ across different insulating materials and a characteristic length scale $\lambda_{tha} = \sqrt{\frac{\hbar}{eB} \left(\frac{\kappa_{xy}}{\kappa_{xx}}\right)}$. 
This length scale $\lambda_{tha}$ represents the correlation between the thermal Hall angle and the mean free path for phonon Hall response. Interestingly, it was discovered that $\lambda_{tha}$ varies only slightly, namely from 2\AA{} to 7\AA{}, among these insulating materials, despite their phonon mean free paths at the respective peak temperatures varying by up to four orders of magnitudes. Motivated by the study of Li et al. \cite{2023NatCo..14.1027L}, we also generated similar plots and examined the range of $\lambda_{tha}$ applicable to our samples. 
Intriguingly, the $\lambda_{tha}$ values for all three samples are between 3\AA{} and 3.5\AA{}, falling within the range of 2\AA{} to 7\AA{} (see Fig.~\ref{FIG.3.}(b)). It is noteworthy that due to the small variation of $\lambda_{tha}$ among different solids, phonon THE has been proposed to be an intrinsic property of insulating solids, as e.g. Y$_2$Ti$_2$O$_7$. However, this scaling behavior holds true even in materials where extrinsic scenarios have been proposed~\cite{2005PhRvL..95o5901S, 2019PhRvB..99m4419H, Ataei_2024, 2023arXiv231010643V5}. 
This aligns with our observation of a weak impact on the thermal Hall ratio from the magnetic Dy$^{3+}$ ions. 
 
To conclude, we conducted a comparative study of thermal transport in Y$_2$Ti$_2$O$_7$, Dy$_2$Ti$_2$O$_7$, and DyYTi$_2$O$_7$ single crystals. The longitudinal thermal conductivities in all three samples are phonon-dominated. The presence of Dy$^{3+}$ magnetic ions significantly affects $\kappa_{xx}$, as seen in the field-dependent data for Dy$_2$Ti$_2$O$_7$ and DyYTi$_2$O$_7$. We measured sizeable field-linear thermal Hall conductivities $\kappa_{xy}$ in all three samples. The temperature dependent $\kappa_{xy}$ show peaks around 15~K, coinciding with the respective peaks of $\kappa_{xx}$. The thermal Hall ratio, $\kappa_{xy}/\kappa_{xx}$, is large (approximately $-10^{-4}$ to $-10^{-3}$)  and aligns with other insulating materials where the THE of phononic origin has been reported \cite{2005PhRvL..95o5901S, 2019PhRvB..99m4419H, 2017NatMa..16..797I, 2020NatPh..16.1108G, 2020NatCo..11.5325B, 2022PhRvX..12b1025L, 2020PhRvL.124j5901L, 2023NatCo..14.1027L, 2022NatCo..13.4604U, 2022PNAS..11908016C, 2024PhRvB.109j4304S, 2024arXiv240402863J, 2023arXiv231010643V5}. 
At low temperature, the thermal Hall ratio of Dy$_2$Ti$_2$O$_7$ is moderately enhanced in comparison to that of Y$_2$Ti$_2$O$_7$, suggesting that the phononic THE is of intrinsic origin for Y$_2$Ti$_2$O$_7$, but can be a combination of both intrinsic and extrinsic origin for Dy$_2$Ti$_2$O$_7$ and DyYTi$_2$O$_7$.

The data of this article are available from Zenodo~\cite{sharma_2024_12755212}.

\begin{acknowledgments}
We thank J. Frielingsdorf and M. Hiertz for the assistance in crystal growth and characterization and G. Grissonnanche for helpful discussions. We acknowledge support by the German Research Foundation via Project No.277146847-CRC1238 (Sub-projects A02 and B01).

\end{acknowledgments}


\begin{thebibliography}{69}%
	\makeatletter
	\providecommand \@ifxundefined [1]{%
		\@ifx{#1\undefined}
	}%
	\providecommand \@ifnum [1]{%
		\ifnum #1\expandafter \@firstoftwo
		\else \expandafter \@secondoftwo
		\fi
	}%
	\providecommand \@ifx [1]{%
		\ifx #1\expandafter \@firstoftwo
		\else \expandafter \@secondoftwo
		\fi
	}%
	\providecommand \natexlab [1]{#1}%
	\providecommand \enquote  [1]{``#1''}%
	\providecommand \bibnamefont  [1]{#1}%
	\providecommand \bibfnamefont [1]{#1}%
	\providecommand \citenamefont [1]{#1}%
	\providecommand \href@noop [0]{\@secondoftwo}%
	\providecommand \href [0]{\begingroup \@sanitize@url \@href}%
	\providecommand \@href[1]{\@@startlink{#1}\@@href}%
	\providecommand \@@href[1]{\endgroup#1\@@endlink}%
	\providecommand \@sanitize@url [0]{\catcode `\\12\catcode `\$12\catcode
		`\&12\catcode `\#12\catcode `\^12\catcode `\_12\catcode `\%12\relax}%
	\providecommand \@@startlink[1]{}%
	\providecommand \@@endlink[0]{}%
	\providecommand \url  [0]{\begingroup\@sanitize@url \@url }%
	\providecommand \@url [1]{\endgroup\@href {#1}{\urlprefix }}%
	\providecommand \urlprefix  [0]{URL }%
	\providecommand \Eprint [0]{\href }%
	\providecommand \doibase [0]{https://doi.org/}%
	\providecommand \selectlanguage [0]{\@gobble}%
	\providecommand \bibinfo  [0]{\@secondoftwo}%
	\providecommand \bibfield  [0]{\@secondoftwo}%
	\providecommand \translation [1]{[#1]}%
	\providecommand \BibitemOpen [0]{}%
	\providecommand \bibitemStop [0]{}%
	\providecommand \bibitemNoStop [0]{.\EOS\space}%
	\providecommand \EOS [0]{\spacefactor3000\relax}%
	\providecommand \BibitemShut  [1]{\csname bibitem#1\endcsname}%
	\let\auto@bib@innerbib\@empty
	\bibitem [{\citenamefont {{Qin}}\ \emph {et~al.}(2012)\citenamefont {{Qin}},
		\citenamefont {{Zhou}},\ and\ \citenamefont {{Shi}}}]{2012PhRvB..86j4305Q}%
	\BibitemOpen
	\bibfield  {author} {\bibinfo {author} {\bibfnamefont {T.}~\bibnamefont
			{{Qin}}}, \bibinfo {author} {\bibfnamefont {J.}~\bibnamefont {{Zhou}}},\ and\
		\bibinfo {author} {\bibfnamefont {J.}~\bibnamefont {{Shi}}},\ }\bibfield
	{title} {\bibinfo {title} {{Berry curvature and the phonon Hall effect}},\
	}\href {https://doi.org/10.1103/PhysRevB.86.104305} {\bibfield  {journal}
		{\bibinfo  {journal} {\prb}\ }\textbf {\bibinfo {volume} {86}},\ \bibinfo
		{eid} {104305} (\bibinfo {year} {2012})}\BibitemShut {NoStop}%
	\bibitem [{\citenamefont {{Saito}}\ \emph {et~al.}(2019)\citenamefont
		{{Saito}}, \citenamefont {{Misaki}}, \citenamefont {{Ishizuka}},\ and\
		\citenamefont {{Nagaosa}}}]{2019PhRvL.123y5901S}%
	\BibitemOpen
	\bibfield  {author} {\bibinfo {author} {\bibfnamefont {T.}~\bibnamefont
			{{Saito}}}, \bibinfo {author} {\bibfnamefont {K.}~\bibnamefont {{Misaki}}},
		\bibinfo {author} {\bibfnamefont {H.}~\bibnamefont {{Ishizuka}}},\ and\
		\bibinfo {author} {\bibfnamefont {N.}~\bibnamefont {{Nagaosa}}},\ }\bibfield
	{title} {\bibinfo {title} {{Berry Phase of Phonons and Thermal Hall Effect in
				Nonmagnetic Insulators}},\ }\href
	{https://doi.org/10.1103/PhysRevLett.123.255901} {\bibfield  {journal}
		{\bibinfo  {journal} {\prl}\ }\textbf {\bibinfo {volume} {123}},\ \bibinfo
		{eid} {255901} (\bibinfo {year} {2019})}\BibitemShut {NoStop}%
	\bibitem [{\citenamefont {{Ideue}}\ \emph {et~al.}(2012)\citenamefont
		{{Ideue}}, \citenamefont {{Onose}}, \citenamefont {{Katsura}}, \citenamefont
		{{Shiomi}}, \citenamefont {{Ishiwata}}, \citenamefont {{Nagaosa}},\ and\
		\citenamefont {{Tokura}}}]{2012PhRvB..85m4411I}%
	\BibitemOpen
	\bibfield  {author} {\bibinfo {author} {\bibfnamefont {T.}~\bibnamefont
			{{Ideue}}}, \bibinfo {author} {\bibfnamefont {Y.}~\bibnamefont {{Onose}}},
		\bibinfo {author} {\bibfnamefont {H.}~\bibnamefont {{Katsura}}}, \bibinfo
		{author} {\bibfnamefont {Y.}~\bibnamefont {{Shiomi}}}, \bibinfo {author}
		{\bibfnamefont {S.}~\bibnamefont {{Ishiwata}}}, \bibinfo {author}
		{\bibfnamefont {N.}~\bibnamefont {{Nagaosa}}},\ and\ \bibinfo {author}
		{\bibfnamefont {Y.}~\bibnamefont {{Tokura}}},\ }\bibfield  {title} {\bibinfo
		{title} {{Effect of lattice geometry on magnon Hall effect in ferromagnetic
				insulators}},\ }\href {https://doi.org/10.1103/PhysRevB.85.134411} {\bibfield
		{journal} {\bibinfo  {journal} {\prb}\ }\textbf {\bibinfo {volume} {85}},\
		\bibinfo {eid} {134411} (\bibinfo {year} {2012})}\BibitemShut {NoStop}%
	\bibitem [{\citenamefont {{Mangeolle}}\ \emph
		{et~al.}(2022{\natexlab{a}})\citenamefont {{Mangeolle}}, \citenamefont
		{{Balents}},\ and\ \citenamefont {{Savary}}}]{2022PhRvB.106x5139M}%
	\BibitemOpen
	\bibfield  {author} {\bibinfo {author} {\bibfnamefont {L.}~\bibnamefont
			{{Mangeolle}}}, \bibinfo {author} {\bibfnamefont {L.}~\bibnamefont
			{{Balents}}},\ and\ \bibinfo {author} {\bibfnamefont {L.}~\bibnamefont
			{{Savary}}},\ }\bibfield  {title} {\bibinfo {title} {{Thermal conductivity
				and theory of inelastic scattering of phonons by collective fluctuations}},\
	}\href {https://doi.org/10.1103/PhysRevB.106.245139} {\bibfield  {journal}
		{\bibinfo  {journal} {\prb}\ }\textbf {\bibinfo {volume} {106}},\ \bibinfo
		{eid} {245139} (\bibinfo {year} {2022}{\natexlab{a}})}\BibitemShut {NoStop}%
	\bibitem [{\citenamefont {{Mangeolle}}\ \emph
		{et~al.}(2022{\natexlab{b}})\citenamefont {{Mangeolle}}, \citenamefont
		{{Balents}},\ and\ \citenamefont {{Savary}}}]{2022PhRvX..12d1031M}%
	\BibitemOpen
	\bibfield  {author} {\bibinfo {author} {\bibfnamefont {L.}~\bibnamefont
			{{Mangeolle}}}, \bibinfo {author} {\bibfnamefont {L.}~\bibnamefont
			{{Balents}}},\ and\ \bibinfo {author} {\bibfnamefont {L.}~\bibnamefont
			{{Savary}}},\ }\bibfield  {title} {\bibinfo {title} {{Phonon Thermal Hall
				Conductivity from Scattering with Collective Fluctuations}},\ }\href
	{https://doi.org/10.1103/PhysRevX.12.041031} {\bibfield  {journal} {\bibinfo
			{journal} {Physical Review X}\ }\textbf {\bibinfo {volume} {12}},\ \bibinfo
		{eid} {041031} (\bibinfo {year} {2022}{\natexlab{b}})}\BibitemShut {NoStop}%
	\bibitem [{\citenamefont {{Zhang}}\ \emph
		{et~al.}(2021{\natexlab{a}})\citenamefont {{Zhang}}, \citenamefont {{Teng}},
		\citenamefont {{Samajdar}}, \citenamefont {{Sachdev}},\ and\ \citenamefont
		{{Scheurer}}}]{2021PhRvB.104c5103Z}%
	\BibitemOpen
	\bibfield  {author} {\bibinfo {author} {\bibfnamefont {Y.}~\bibnamefont
			{{Zhang}}}, \bibinfo {author} {\bibfnamefont {Y.}~\bibnamefont {{Teng}}},
		\bibinfo {author} {\bibfnamefont {R.}~\bibnamefont {{Samajdar}}}, \bibinfo
		{author} {\bibfnamefont {S.}~\bibnamefont {{Sachdev}}},\ and\ \bibinfo
		{author} {\bibfnamefont {M.~S.}\ \bibnamefont {{Scheurer}}},\ }\bibfield
	{title} {\bibinfo {title} {{Phonon Hall viscosity from phonon-spinon
				interactions}},\ }\href {https://doi.org/10.1103/PhysRevB.104.035103}
	{\bibfield  {journal} {\bibinfo  {journal} {\prb}\ }\textbf {\bibinfo
			{volume} {104}},\ \bibinfo {eid} {035103} (\bibinfo {year}
		{2021}{\natexlab{a}})}\BibitemShut {NoStop}%
	\bibitem [{\citenamefont {{Zhang}}\ \emph {et~al.}(2019)\citenamefont
		{{Zhang}}, \citenamefont {{Zhang}}, \citenamefont {{Okamoto}},\ and\
		\citenamefont {{Xiao}}}]{2019PhRvL.123p7202Z}%
	\BibitemOpen
	\bibfield  {author} {\bibinfo {author} {\bibfnamefont {X.}~\bibnamefont
			{{Zhang}}}, \bibinfo {author} {\bibfnamefont {Y.}~\bibnamefont {{Zhang}}},
		\bibinfo {author} {\bibfnamefont {S.}~\bibnamefont {{Okamoto}}},\ and\
		\bibinfo {author} {\bibfnamefont {D.}~\bibnamefont {{Xiao}}},\ }\bibfield
	{title} {\bibinfo {title} {{Thermal Hall Effect Induced by Magnon-Phonon
				Interactions}},\ }\href {https://doi.org/10.1103/PhysRevLett.123.167202}
	{\bibfield  {journal} {\bibinfo  {journal} {\prl}\ }\textbf {\bibinfo
			{volume} {123}},\ \bibinfo {eid} {167202} (\bibinfo {year}
		{2019})}\BibitemShut {NoStop}%
	\bibitem [{\citenamefont {{Sheng}}\ \emph {et~al.}(2006)\citenamefont
		{{Sheng}}, \citenamefont {{Sheng}},\ and\ \citenamefont
		{{Ting}}}]{2006PhRvL..96o5901S}%
	\BibitemOpen
	\bibfield  {author} {\bibinfo {author} {\bibfnamefont {L.}~\bibnamefont
			{{Sheng}}}, \bibinfo {author} {\bibfnamefont {D.~N.}\ \bibnamefont
			{{Sheng}}},\ and\ \bibinfo {author} {\bibfnamefont {C.~S.}\ \bibnamefont
			{{Ting}}},\ }\bibfield  {title} {\bibinfo {title} {{Theory of the Phonon Hall
				Effect in Paramagnetic Dielectrics}},\ }\href
	{https://doi.org/10.1103/PhysRevLett.96.155901} {\bibfield  {journal}
		{\bibinfo  {journal} {\prl}\ }\textbf {\bibinfo {volume} {96}},\ \bibinfo
		{eid} {155901} (\bibinfo {year} {2006})}\BibitemShut {NoStop}%
	\bibitem [{\citenamefont {{Takahashi}}\ and\ \citenamefont
		{{Nagaosa}}(2016)}]{2016PhRvL.117u7205T}%
	\BibitemOpen
	\bibfield  {author} {\bibinfo {author} {\bibfnamefont {R.}~\bibnamefont
			{{Takahashi}}}\ and\ \bibinfo {author} {\bibfnamefont {N.}~\bibnamefont
			{{Nagaosa}}},\ }\bibfield  {title} {\bibinfo {title} {{Berry Curvature in
				Magnon-Phonon Hybrid Systems}},\ }\href
	{https://doi.org/10.1103/PhysRevLett.117.217205} {\bibfield  {journal}
		{\bibinfo  {journal} {\prl}\ }\textbf {\bibinfo {volume} {117}},\ \bibinfo
		{eid} {217205} (\bibinfo {year} {2016})}\BibitemShut {NoStop}%
	\bibitem [{\citenamefont {{Guo}}\ \emph {et~al.}(2022)\citenamefont {{Guo}},
		\citenamefont {{Joshi}},\ and\ \citenamefont
		{{Sachdev}}}]{2022PNAS..11915141G}%
	\BibitemOpen
	\bibfield  {author} {\bibinfo {author} {\bibfnamefont {H.}~\bibnamefont
			{{Guo}}}, \bibinfo {author} {\bibfnamefont {D.~G.}\ \bibnamefont {{Joshi}}},\
		and\ \bibinfo {author} {\bibfnamefont {S.}~\bibnamefont {{Sachdev}}},\
	}\bibfield  {title} {\bibinfo {title} {{Resonant thermal Hall effect of
				phonons coupled to dynamical defects}},\ }\href
	{https://doi.org/10.1073/pnas.2215141119} {\bibfield  {journal} {\bibinfo
			{journal} {Proceedings of the National Academy of Science}\ }\textbf
		{\bibinfo {volume} {119}},\ \bibinfo {eid} {e2215141119} (\bibinfo {year}
		{2022})}\BibitemShut {NoStop}%
	\bibitem [{\citenamefont {{Guo}}\ and\ \citenamefont
		{{Sachdev}}(2021)}]{2021PhRvB.103t5115G}%
	\BibitemOpen
	\bibfield  {author} {\bibinfo {author} {\bibfnamefont {H.}~\bibnamefont
			{{Guo}}}\ and\ \bibinfo {author} {\bibfnamefont {S.}~\bibnamefont
			{{Sachdev}}},\ }\bibfield  {title} {\bibinfo {title} {{Extrinsic phonon
				thermal Hall transport from Hall viscosity}},\ }\href
	{https://doi.org/10.1103/PhysRevB.103.205115} {\bibfield  {journal} {\bibinfo
			{journal} {\prb}\ }\textbf {\bibinfo {volume} {103}},\ \bibinfo {eid}
		{205115} (\bibinfo {year} {2021})}\BibitemShut {NoStop}%
\bibitem [{\citenamefont {Guo}(2023)}]{PhysRevResearch.5.033197}
\BibitemOpen
\bibfield  {author} {\bibinfo {author} {\bibfnamefont {Haoyu}\ \bibnamefont {Guo}},\ }\bibfield  {title} {\bibinfo {title} {Phonon thermal hall effect in a non-Kramers paramagnet},\ }\href {https://link.aps.org/doi/10.1103/PhysRevResearch.5.033197} {\bibfield  {journal} {\bibinfo  {journal} {Phys. Rev. Res.}\ }\textbf {\bibinfo {volume} {5}},\ \bibinfo {pages} {033197} (\bibinfo {year} {2023})}\BibitemShut {NoStop}%
	\bibitem [{\citenamefont {{Flebus}}\ and\ \citenamefont
		{{MacDonald}}(2022)}]{2022PhRvB.105v0301F}%
	\BibitemOpen
	\bibfield  {author} {\bibinfo {author} {\bibfnamefont {B.}~\bibnamefont
			{{Flebus}}}\ and\ \bibinfo {author} {\bibfnamefont {A.~H.}\ \bibnamefont
			{{MacDonald}}},\ }\bibfield  {title} {\bibinfo {title} {{Charged defects and
				phonon Hall effects in ionic crystals}},\ }\href
	{https://doi.org/10.1103/PhysRevB.105.L220301} {\bibfield  {journal}
		{\bibinfo  {journal} {\prb}\ }\textbf {\bibinfo {volume} {105}},\ \bibinfo
		{eid} {L220301} (\bibinfo {year} {2022})}\BibitemShut {NoStop}%
	\bibitem [{\citenamefont {{Chen}}\ \emph {et~al.}(2020)\citenamefont {{Chen}},
		\citenamefont {{Kivelson}},\ and\ \citenamefont
		{{Sun}}}]{2020PhRvL.124p7601C}%
	\BibitemOpen
	\bibfield  {author} {\bibinfo {author} {\bibfnamefont {J.-Y.}\ \bibnamefont
			{{Chen}}}, \bibinfo {author} {\bibfnamefont {S.~A.}\ \bibnamefont
			{{Kivelson}}},\ and\ \bibinfo {author} {\bibfnamefont {X.-Q.}\ \bibnamefont
			{{Sun}}},\ }\bibfield  {title} {\bibinfo {title} {{Enhanced Thermal Hall
				Effect in Nearly Ferroelectric Insulators}},\ }\href
	{https://doi.org/10.1103/PhysRevLett.124.167601} {\bibfield  {journal}
		{\bibinfo  {journal} {\prl}\ }\textbf {\bibinfo {volume} {124}},\ \bibinfo
		{eid} {167601} (\bibinfo {year} {2020})}\BibitemShut {NoStop}%
	\bibitem [{\citenamefont {{Sun}}\ \emph {et~al.}(2022)\citenamefont {{Sun}},
		\citenamefont {{Chen}},\ and\ \citenamefont
		{{Kivelson}}}]{2022PhRvB.106n4111S}%
	\BibitemOpen
	\bibfield  {author} {\bibinfo {author} {\bibfnamefont {X.-Q.}\ \bibnamefont
			{{Sun}}}, \bibinfo {author} {\bibfnamefont {J.-Y.}\ \bibnamefont {{Chen}}},\
		and\ \bibinfo {author} {\bibfnamefont {S.~A.}\ \bibnamefont {{Kivelson}}},\
	}\bibfield  {title} {\bibinfo {title} {{Large extrinsic phonon thermal Hall
				effect from resonant scattering}},\ }\href
	{https://doi.org/10.1103/PhysRevB.106.144111} {\bibfield  {journal} {\bibinfo
			{journal} {\prb}\ }\textbf {\bibinfo {volume} {106}},\ \bibinfo {eid}
		{144111} (\bibinfo {year} {2022})}\BibitemShut {NoStop}%
	\bibitem [{\citenamefont {{Mori}}\ \emph {et~al.}(2014)\citenamefont {{Mori}},
		\citenamefont {{Spencer-Smith}}, \citenamefont {{Sushkov}},\ and\
		\citenamefont {{Maekawa}}}]{2014PhRvL.113z5901M}%
	\BibitemOpen
	\bibfield  {author} {\bibinfo {author} {\bibfnamefont {M.}~\bibnamefont
			{{Mori}}}, \bibinfo {author} {\bibfnamefont {A.}~\bibnamefont
			{{Spencer-Smith}}}, \bibinfo {author} {\bibfnamefont {O.~P.}\ \bibnamefont
			{{Sushkov}}},\ and\ \bibinfo {author} {\bibfnamefont {S.}~\bibnamefont
			{{Maekawa}}},\ }\bibfield  {title} {\bibinfo {title} {{Origin of the Phonon
				Hall Effect in Rare-Earth Garnets}},\ }\href
	{https://doi.org/10.1103/PhysRevLett.113.265901} {\bibfield  {journal}
		{\bibinfo  {journal} {\prl}\ }\textbf {\bibinfo {volume} {113}},\ \bibinfo
		{eid} {265901} (\bibinfo {year} {2014})}\BibitemShut {NoStop}%
	\bibitem [{\citenamefont {{Strohm}}\ \emph {et~al.}(2005)\citenamefont
		{{Strohm}}, \citenamefont {{Rikken}},\ and\ \citenamefont
		{{Wyder}}}]{2005PhRvL..95o5901S}%
	\BibitemOpen
	\bibfield  {author} {\bibinfo {author} {\bibfnamefont {C.}~\bibnamefont
			{{Strohm}}}, \bibinfo {author} {\bibfnamefont {G.~L.~J.~A.}\ \bibnamefont
			{{Rikken}}},\ and\ \bibinfo {author} {\bibfnamefont {P.}~\bibnamefont
			{{Wyder}}},\ }\bibfield  {title} {\bibinfo {title} {{Phenomenological
				Evidence for the Phonon Hall Effect}},\ }\href
	{https://doi.org/10.1103/PhysRevLett.95.155901} {\bibfield  {journal}
		{\bibinfo  {journal} {\prl}\ }\textbf {\bibinfo {volume} {95}},\ \bibinfo
		{eid} {155901} (\bibinfo {year} {2005})}\BibitemShut {NoStop}%
	\bibitem [{\citenamefont {{Grissonnanche}}\ \emph {et~al.}(2019)\citenamefont
		{{Grissonnanche}}, \citenamefont {{Legros}}, \citenamefont {{Badoux}},
		\citenamefont {{Lefran{\c{c}}ois}}, \citenamefont {{Zatko}}, \citenamefont
		{{Lizaire}}, \citenamefont {{Lalibert{\'e}}}, \citenamefont {{Gourgout}},
		\citenamefont {{Zhou}}, \citenamefont {{Pyon}}, \citenamefont {{Takayama}},
		\citenamefont {{Takagi}}, \citenamefont {{Ono}}, \citenamefont
		{{Doiron-Leyraud}},\ and\ \citenamefont {{Taillefer}}}]{2019Natur.571..376G}%
	\BibitemOpen
	\bibfield  {author} {\bibinfo {author} {\bibfnamefont {G.}~\bibnamefont
			{{Grissonnanche}}}, \bibinfo {author} {\bibfnamefont {A.}~\bibnamefont
			{{Legros}}}, \bibinfo {author} {\bibfnamefont {S.}~\bibnamefont {{Badoux}}},
		\bibinfo {author} {\bibfnamefont {E.}~\bibnamefont {{Lefran{\c{c}}ois}}},
		\bibinfo {author} {\bibfnamefont {V.}~\bibnamefont {{Zatko}}}, \bibinfo
		{author} {\bibfnamefont {M.}~\bibnamefont {{Lizaire}}}, \bibinfo {author}
		{\bibfnamefont {F.}~\bibnamefont {{Lalibert{\'e}}}}, \bibinfo {author}
		{\bibfnamefont {A.}~\bibnamefont {{Gourgout}}}, \bibinfo {author}
		{\bibfnamefont {J.~S.}\ \bibnamefont {{Zhou}}}, \bibinfo {author}
		{\bibfnamefont {S.}~\bibnamefont {{Pyon}}}, \bibinfo {author} {\bibfnamefont
			{T.}~\bibnamefont {{Takayama}}}, \bibinfo {author} {\bibfnamefont
			{H.}~\bibnamefont {{Takagi}}}, \bibinfo {author} {\bibfnamefont
			{S.}~\bibnamefont {{Ono}}}, \bibinfo {author} {\bibfnamefont
			{N.}~\bibnamefont {{Doiron-Leyraud}}},\ and\ \bibinfo {author} {\bibfnamefont
			{L.}~\bibnamefont {{Taillefer}}},\ }\bibfield  {title} {\bibinfo {title}
		{{Giant thermal Hall conductivity in the pseudogap phase of cuprate
				superconductors}},\ }\href {https://doi.org/10.1038/s41586-019-1375-0}
	{\bibfield  {journal} {\bibinfo  {journal} {\nat}\ }\textbf {\bibinfo
			{volume} {571}},\ \bibinfo {pages} {376} (\bibinfo {year}
		{2019})}\BibitemShut {NoStop}%
	\bibitem [{\citenamefont {{Grissonnanche}}\ \emph {et~al.}(2020)\citenamefont
		{{Grissonnanche}}, \citenamefont {{Th{\'e}riault}}, \citenamefont
		{{Gourgout}}, \citenamefont {{Boulanger}}, \citenamefont
		{{Lefran{\c{c}}ois}}, \citenamefont {{Ataei}}, \citenamefont
		{{Lalibert{\'e}}}, \citenamefont {{Dion}}, \citenamefont {{Zhou}},
		\citenamefont {{Pyon}}, \citenamefont {{Takayama}}, \citenamefont {{Takagi}},
		\citenamefont {{Doiron-Leyraud}},\ and\ \citenamefont
		{{Taillefer}}}]{2020NatPh..16.1108G}%
	\BibitemOpen
	\bibfield  {author} {\bibinfo {author} {\bibfnamefont {G.}~\bibnamefont
			{{Grissonnanche}}}, \bibinfo {author} {\bibfnamefont {S.}~\bibnamefont
			{{Th{\'e}riault}}}, \bibinfo {author} {\bibfnamefont {A.}~\bibnamefont
			{{Gourgout}}}, \bibinfo {author} {\bibfnamefont {M.~E.}\ \bibnamefont
			{{Boulanger}}}, \bibinfo {author} {\bibfnamefont {E.}~\bibnamefont
			{{Lefran{\c{c}}ois}}}, \bibinfo {author} {\bibfnamefont {A.}~\bibnamefont
			{{Ataei}}}, \bibinfo {author} {\bibfnamefont {F.}~\bibnamefont
			{{Lalibert{\'e}}}}, \bibinfo {author} {\bibfnamefont {M.}~\bibnamefont
			{{Dion}}}, \bibinfo {author} {\bibfnamefont {J.~S.}\ \bibnamefont {{Zhou}}},
		\bibinfo {author} {\bibfnamefont {S.}~\bibnamefont {{Pyon}}}, \bibinfo
		{author} {\bibfnamefont {T.}~\bibnamefont {{Takayama}}}, \bibinfo {author}
		{\bibfnamefont {H.}~\bibnamefont {{Takagi}}}, \bibinfo {author}
		{\bibfnamefont {N.}~\bibnamefont {{Doiron-Leyraud}}},\ and\ \bibinfo {author}
		{\bibfnamefont {L.}~\bibnamefont {{Taillefer}}},\ }\bibfield  {title}
	{\bibinfo {title} {{Chiral phonons in the pseudogap phase of cuprates}},\
	}\href {https://doi.org/10.1038/s41567-020-0965-y} {\bibfield  {journal}
		{\bibinfo  {journal} {Nature Physics}\ }\textbf {\bibinfo {volume} {16}},\
		\bibinfo {pages} {1108} (\bibinfo {year} {2020})}\BibitemShut {NoStop}%
	\bibitem [{\citenamefont {{Chen}}\ \emph {et~al.}(2022)\citenamefont {{Chen}},
		\citenamefont {{Boulanger}}, \citenamefont {{Wang}}, \citenamefont
		{{Tafti}},\ and\ \citenamefont {{Taillefer}}}]{2022PNAS..11908016C}%
	\BibitemOpen
	\bibfield  {author} {\bibinfo {author} {\bibfnamefont {L.}~\bibnamefont
			{{Chen}}}, \bibinfo {author} {\bibfnamefont {M.-E.}\ \bibnamefont
			{{Boulanger}}}, \bibinfo {author} {\bibfnamefont {Z.-C.}\ \bibnamefont
			{{Wang}}}, \bibinfo {author} {\bibfnamefont {F.}~\bibnamefont {{Tafti}}},\
		and\ \bibinfo {author} {\bibfnamefont {L.}~\bibnamefont {{Taillefer}}},\
	}\bibfield  {title} {\bibinfo {title} {{Large phonon thermal Hall
				conductivity in the antiferromagnetic insulator Cu$_{3}$TeO$_{6}$}},\ }\href
	{https://doi.org/10.1073/pnas.2208016119} {\bibfield  {journal} {\bibinfo
			{journal} {Proceedings of the National Academy of Science}\ }\textbf
		{\bibinfo {volume} {119}},\ \bibinfo {eid} {e2208016119} (\bibinfo {year}
		{2022})}\BibitemShut {NoStop}%
	\bibitem [{\citenamefont {{Boulanger}}\ \emph {et~al.}(2020)\citenamefont
		{{Boulanger}}, \citenamefont {{Grissonnanche}}, \citenamefont {{Badoux}},
		\citenamefont {{Allaire}}, \citenamefont {{Lefran{\c{c}}ois}}, \citenamefont
		{{Legros}}, \citenamefont {{Gourgout}}, \citenamefont {{Dion}}, \citenamefont
		{{Wang}}, \citenamefont {{Chen}}, \citenamefont {{Liang}}, \citenamefont
		{{Hardy}}, \citenamefont {{Bonn}},\ and\ \citenamefont
		{{Taillefer}}}]{2020NatCo..11.5325B}%
	\BibitemOpen
	\bibfield  {author} {\bibinfo {author} {\bibfnamefont {M.-E.}\ \bibnamefont
			{{Boulanger}}}, \bibinfo {author} {\bibfnamefont {G.}~\bibnamefont
			{{Grissonnanche}}}, \bibinfo {author} {\bibfnamefont {S.}~\bibnamefont
			{{Badoux}}}, \bibinfo {author} {\bibfnamefont {A.}~\bibnamefont {{Allaire}}},
		\bibinfo {author} {\bibfnamefont {{\'E}.}~\bibnamefont {{Lefran{\c{c}}ois}}},
		\bibinfo {author} {\bibfnamefont {A.}~\bibnamefont {{Legros}}}, \bibinfo
		{author} {\bibfnamefont {A.}~\bibnamefont {{Gourgout}}}, \bibinfo {author}
		{\bibfnamefont {M.}~\bibnamefont {{Dion}}}, \bibinfo {author} {\bibfnamefont
			{C.~H.}\ \bibnamefont {{Wang}}}, \bibinfo {author} {\bibfnamefont {X.~H.}\
			\bibnamefont {{Chen}}}, \bibinfo {author} {\bibfnamefont {R.}~\bibnamefont
			{{Liang}}}, \bibinfo {author} {\bibfnamefont {W.~N.}\ \bibnamefont
			{{Hardy}}}, \bibinfo {author} {\bibfnamefont {D.~A.}\ \bibnamefont
			{{Bonn}}},\ and\ \bibinfo {author} {\bibfnamefont {L.}~\bibnamefont
			{{Taillefer}}},\ }\bibfield  {title} {\bibinfo {title} {{Thermal Hall
				conductivity in the cuprate Mott insulators Nd$_{2}$CuO$_{4}$ and
				Sr$_{2}$CuO$_{2}$Cl$_{2}$}},\ }\href
	{https://doi.org/10.1038/s41467-020-18881-z} {\bibfield  {journal} {\bibinfo
			{journal} {Nature Communications}\ }\textbf {\bibinfo {volume} {11}},\
		\bibinfo {eid} {5325} (\bibinfo {year} {2020})}\BibitemShut {NoStop}%
	\bibitem [{\citenamefont {{Sugii}}\ \emph {et~al.}(2017)\citenamefont
		{{Sugii}}, \citenamefont {{Shimozawa}}, \citenamefont {{Watanabe}},
		\citenamefont {{Suzuki}}, \citenamefont {{Halim}}, \citenamefont {{Kimata}},
		\citenamefont {{Matsumoto}}, \citenamefont {{Nakatsuji}},\ and\ \citenamefont
		{{Yamashita}}}]{2017PhRvL.118n5902S}%
	\BibitemOpen
	\bibfield  {author} {\bibinfo {author} {\bibfnamefont {K.}~\bibnamefont
			{{Sugii}}}, \bibinfo {author} {\bibfnamefont {M.}~\bibnamefont
			{{Shimozawa}}}, \bibinfo {author} {\bibfnamefont {D.}~\bibnamefont
			{{Watanabe}}}, \bibinfo {author} {\bibfnamefont {Y.}~\bibnamefont
			{{Suzuki}}}, \bibinfo {author} {\bibfnamefont {M.}~\bibnamefont {{Halim}}},
		\bibinfo {author} {\bibfnamefont {M.}~\bibnamefont {{Kimata}}}, \bibinfo
		{author} {\bibfnamefont {Y.}~\bibnamefont {{Matsumoto}}}, \bibinfo {author}
		{\bibfnamefont {S.}~\bibnamefont {{Nakatsuji}}},\ and\ \bibinfo {author}
		{\bibfnamefont {M.}~\bibnamefont {{Yamashita}}},\ }\bibfield  {title}
	{\bibinfo {title} {{Thermal Hall Effect in a Phonon-Glass
				Ba$_{3}$CuSb$_{2}$O$_{9}$}},\ }\href
	{https://doi.org/10.1103/PhysRevLett.118.145902} {\bibfield  {journal}
		{\bibinfo  {journal} {\prl}\ }\textbf {\bibinfo {volume} {118}},\ \bibinfo
		{eid} {145902} (\bibinfo {year} {2017})}\BibitemShut {NoStop}%
	\bibitem [{\citenamefont {{Ideue}}\ \emph {et~al.}(2017)\citenamefont
		{{Ideue}}, \citenamefont {{Kurumaji}}, \citenamefont {{Ishiwata}},\ and\
		\citenamefont {{Tokura}}}]{2017NatMa..16..797I}%
	\BibitemOpen
	\bibfield  {author} {\bibinfo {author} {\bibfnamefont {T.}~\bibnamefont
			{{Ideue}}}, \bibinfo {author} {\bibfnamefont {T.}~\bibnamefont {{Kurumaji}}},
		\bibinfo {author} {\bibfnamefont {S.}~\bibnamefont {{Ishiwata}}},\ and\
		\bibinfo {author} {\bibfnamefont {Y.}~\bibnamefont {{Tokura}}},\ }\bibfield
	{title} {\bibinfo {title} {{Giant thermal Hall effect in multiferroics}},\
	}\href {https://doi.org/10.1038/nmat4905} {\bibfield  {journal} {\bibinfo
			{journal} {Nature Materials}\ }\textbf {\bibinfo {volume} {16}},\ \bibinfo
		{pages} {797} (\bibinfo {year} {2017})}\BibitemShut {NoStop}%
	\bibitem [{\citenamefont {{Uehara}}\ \emph {et~al.}(2022)\citenamefont
		{{Uehara}}, \citenamefont {{Ohtsuki}}, \citenamefont {{Udagawa}},
		\citenamefont {{Nakatsuji}},\ and\ \citenamefont
		{{Machida}}}]{2022NatCo..13.4604U}%
	\BibitemOpen
	\bibfield  {author} {\bibinfo {author} {\bibfnamefont {T.}~\bibnamefont
			{{Uehara}}}, \bibinfo {author} {\bibfnamefont {T.}~\bibnamefont {{Ohtsuki}}},
		\bibinfo {author} {\bibfnamefont {M.}~\bibnamefont {{Udagawa}}}, \bibinfo
		{author} {\bibfnamefont {S.}~\bibnamefont {{Nakatsuji}}},\ and\ \bibinfo
		{author} {\bibfnamefont {Y.}~\bibnamefont {{Machida}}},\ }\bibfield  {title}
	{\bibinfo {title} {{Phonon thermal Hall effect in a metallic spin ice}},\
	}\href {https://doi.org/10.1038/s41467-022-32375-0} {\bibfield  {journal}
		{\bibinfo  {journal} {Nature Communications}\ }\textbf {\bibinfo {volume}
			{13}},\ \bibinfo {eid} {4604} (\bibinfo {year} {2022})}\BibitemShut {NoStop}%
	\bibitem [{\citenamefont {{Hirschberger}}\ \emph
		{et~al.}(2015{\natexlab{a}})\citenamefont {{Hirschberger}}, \citenamefont
		{{Krizan}}, \citenamefont {{Cava}},\ and\ \citenamefont
		{{Ong}}}]{2015Sci...348..106H}%
	\BibitemOpen
	\bibfield  {author} {\bibinfo {author} {\bibfnamefont {M.}~\bibnamefont
			{{Hirschberger}}}, \bibinfo {author} {\bibfnamefont {J.~W.}\ \bibnamefont
			{{Krizan}}}, \bibinfo {author} {\bibfnamefont {R.~J.}\ \bibnamefont
			{{Cava}}},\ and\ \bibinfo {author} {\bibfnamefont {N.~P.}\ \bibnamefont
			{{Ong}}},\ }\bibfield  {title} {\bibinfo {title} {{Large thermal Hall
				conductivity of neutral spin excitations in a frustrated quantum magnet}},\
	}\href {https://doi.org/10.1126/science.1257340} {\bibfield  {journal}
		{\bibinfo  {journal} {Science}\ }\textbf {\bibinfo {volume} {348}},\ \bibinfo
		{pages} {106} (\bibinfo {year} {2015}{\natexlab{a}})}\BibitemShut {NoStop}%
	\bibitem [{\citenamefont {{Hirokane}}\ \emph {et~al.}(2019)\citenamefont
		{{Hirokane}}, \citenamefont {{Nii}}, \citenamefont {{Tomioka}},\ and\
		\citenamefont {{Onose}}}]{2019PhRvB..99m4419H}%
	\BibitemOpen
	\bibfield  {author} {\bibinfo {author} {\bibfnamefont {Y.}~\bibnamefont
			{{Hirokane}}}, \bibinfo {author} {\bibfnamefont {Y.}~\bibnamefont {{Nii}}},
		\bibinfo {author} {\bibfnamefont {Y.}~\bibnamefont {{Tomioka}}},\ and\
		\bibinfo {author} {\bibfnamefont {Y.}~\bibnamefont {{Onose}}},\ }\bibfield
	{title} {\bibinfo {title} {{Phononic thermal Hall effect in diluted terbium
				oxides}},\ }\href {https://doi.org/10.1103/PhysRevB.99.134419} {\bibfield
		{journal} {\bibinfo  {journal} {\prb}\ }\textbf {\bibinfo {volume} {99}},\
		\bibinfo {eid} {134419} (\bibinfo {year} {2019})}\BibitemShut {NoStop}%
	\bibitem [{\citenamefont {{Sim}}\ \emph {et~al.}(2021)\citenamefont {{Sim}},
		\citenamefont {{Yang}}, \citenamefont {{Kim}}, \citenamefont {{Coak}},
		\citenamefont {{Itoh}}, \citenamefont {{Noda}},\ and\ \citenamefont
		{{Park}}}]{2021PhRvL.126a5901S}%
	\BibitemOpen
	\bibfield  {author} {\bibinfo {author} {\bibfnamefont {S.}~\bibnamefont
			{{Sim}}}, \bibinfo {author} {\bibfnamefont {H.}~\bibnamefont {{Yang}}},
		\bibinfo {author} {\bibfnamefont {H.-L.}\ \bibnamefont {{Kim}}}, \bibinfo
		{author} {\bibfnamefont {M.~J.}\ \bibnamefont {{Coak}}}, \bibinfo {author}
		{\bibfnamefont {M.}~\bibnamefont {{Itoh}}}, \bibinfo {author} {\bibfnamefont
			{Y.}~\bibnamefont {{Noda}}},\ and\ \bibinfo {author} {\bibfnamefont {J.-G.}\
			\bibnamefont {{Park}}},\ }\bibfield  {title} {\bibinfo {title} {{Sizable
				Suppression of Thermal Hall Effect upon Isotopic Substitution in
				SrTiO$_{3}$}},\ }\href {https://doi.org/10.1103/PhysRevLett.126.015901}
	{\bibfield  {journal} {\bibinfo  {journal} {\prl}\ }\textbf {\bibinfo
			{volume} {126}},\ \bibinfo {eid} {015901} (\bibinfo {year}
		{2021})}\BibitemShut {NoStop}%
	\bibitem [{\citenamefont {{Hirschberger}}\ \emph
		{et~al.}(2015{\natexlab{b}})\citenamefont {{Hirschberger}}, \citenamefont
		{{Chisnell}}, \citenamefont {{Lee}},\ and\ \citenamefont
		{{Ong}}}]{2015PhRvL.115j6603H}%
	\BibitemOpen
	\bibfield  {author} {\bibinfo {author} {\bibfnamefont {M.}~\bibnamefont
			{{Hirschberger}}}, \bibinfo {author} {\bibfnamefont {R.}~\bibnamefont
			{{Chisnell}}}, \bibinfo {author} {\bibfnamefont {Y.~S.}\ \bibnamefont
			{{Lee}}},\ and\ \bibinfo {author} {\bibfnamefont {N.~P.}\ \bibnamefont
			{{Ong}}},\ }\bibfield  {title} {\bibinfo {title} {{Thermal Hall Effect of
				Spin Excitations in a Kagome Magnet}},\ }\href
	{https://doi.org/10.1103/PhysRevLett.115.106603} {\bibfield  {journal}
		{\bibinfo  {journal} {\prl}\ }\textbf {\bibinfo {volume} {115}},\ \bibinfo
		{eid} {106603} (\bibinfo {year} {2015}{\natexlab{b}})}\BibitemShut {NoStop}%
	\bibitem [{\citenamefont {{Kim}}\ \emph {et~al.}(2024)\citenamefont {{Kim}},
		\citenamefont {{Saito}}, \citenamefont {{Yang}}, \citenamefont {{Ishizuka}},
		\citenamefont {{Coak}}, \citenamefont {{Lee}}, \citenamefont {{Sim}},
		\citenamefont {{Oh}}, \citenamefont {{Nagaosa}},\ and\ \citenamefont
		{{Park}}}]{2024NatCo..15..243K}%
	\BibitemOpen
	\bibfield  {author} {\bibinfo {author} {\bibfnamefont {H.-L.}\ \bibnamefont
			{{Kim}}}, \bibinfo {author} {\bibfnamefont {T.}~\bibnamefont {{Saito}}},
		\bibinfo {author} {\bibfnamefont {H.}~\bibnamefont {{Yang}}}, \bibinfo
		{author} {\bibfnamefont {H.}~\bibnamefont {{Ishizuka}}}, \bibinfo {author}
		{\bibfnamefont {M.~J.}\ \bibnamefont {{Coak}}}, \bibinfo {author}
		{\bibfnamefont {J.~H.}\ \bibnamefont {{Lee}}}, \bibinfo {author}
		{\bibfnamefont {H.}~\bibnamefont {{Sim}}}, \bibinfo {author} {\bibfnamefont
			{Y.~S.}\ \bibnamefont {{Oh}}}, \bibinfo {author} {\bibfnamefont
			{N.}~\bibnamefont {{Nagaosa}}},\ and\ \bibinfo {author} {\bibfnamefont
			{J.-G.}\ \bibnamefont {{Park}}},\ }\bibfield  {title} {\bibinfo {title}
		{{Thermal Hall effects due to topological spin fluctuations in YMnO$_{3}$}},\
	}\href {https://doi.org/10.1038/s41467-023-44448-9} {\bibfield  {journal}
		{\bibinfo  {journal} {Nature Communications}\ }\textbf {\bibinfo {volume}
			{15}},\ \bibinfo {eid} {243} (\bibinfo {year} {2024})}\BibitemShut {NoStop}%
	\bibitem [{\citenamefont {{Hentrich}}\ \emph {et~al.}(2019)\citenamefont
		{{Hentrich}}, \citenamefont {{Roslova}}, \citenamefont {{Isaeva}},
		\citenamefont {{Doert}}, \citenamefont {{Brenig}}, \citenamefont
		{{B{\"u}chner}},\ and\ \citenamefont {{Hess}}}]{2019PhRvB..99h5136H}%
	\BibitemOpen
	\bibfield  {author} {\bibinfo {author} {\bibfnamefont {R.}~\bibnamefont
			{{Hentrich}}}, \bibinfo {author} {\bibfnamefont {M.}~\bibnamefont
			{{Roslova}}}, \bibinfo {author} {\bibfnamefont {A.}~\bibnamefont {{Isaeva}}},
		\bibinfo {author} {\bibfnamefont {T.}~\bibnamefont {{Doert}}}, \bibinfo
		{author} {\bibfnamefont {W.}~\bibnamefont {{Brenig}}}, \bibinfo {author}
		{\bibfnamefont {B.}~\bibnamefont {{B{\"u}chner}}},\ and\ \bibinfo {author}
		{\bibfnamefont {C.}~\bibnamefont {{Hess}}},\ }\bibfield  {title} {\bibinfo
		{title} {{Large thermal hall effect in {\ensuremath{\alpha}} -RuCl$_{3}$ :
				Evidence for heat transport by Kitaev-Heisenberg paramagnons}},\ }\href
	{https://doi.org/10.1103/PhysRevB.99.085136} {\bibfield  {journal} {\bibinfo
			{journal} {\prb}\ }\textbf {\bibinfo {volume} {99}},\ \bibinfo {eid} {085136}
		(\bibinfo {year} {2019})}\BibitemShut {NoStop}%
	\bibitem [{\citenamefont {{Zhang}}\ \emph
		{et~al.}(2021{\natexlab{b}})\citenamefont {{Zhang}}, \citenamefont {{Xu}},
		\citenamefont {{Carnahan}}, \citenamefont {{Sretenovic}}, \citenamefont
		{{Suri}}, \citenamefont {{Xiao}},\ and\ \citenamefont
		{{Ke}}}]{2021PhRvL.127x7202Z}%
	\BibitemOpen
	\bibfield  {author} {\bibinfo {author} {\bibfnamefont {H.}~\bibnamefont
			{{Zhang}}}, \bibinfo {author} {\bibfnamefont {C.}~\bibnamefont {{Xu}}},
		\bibinfo {author} {\bibfnamefont {C.}~\bibnamefont {{Carnahan}}}, \bibinfo
		{author} {\bibfnamefont {M.}~\bibnamefont {{Sretenovic}}}, \bibinfo {author}
		{\bibfnamefont {N.}~\bibnamefont {{Suri}}}, \bibinfo {author} {\bibfnamefont
			{D.}~\bibnamefont {{Xiao}}},\ and\ \bibinfo {author} {\bibfnamefont
			{X.}~\bibnamefont {{Ke}}},\ }\bibfield  {title} {\bibinfo {title} {{Anomalous
				Thermal Hall Effect in an Insulating van der Waals Magnet}},\ }\href
	{https://doi.org/10.1103/PhysRevLett.127.247202} {\bibfield  {journal}
		{\bibinfo  {journal} {\prl}\ }\textbf {\bibinfo {volume} {127}},\ \bibinfo
		{eid} {247202} (\bibinfo {year} {2021}{\natexlab{b}})}\BibitemShut {NoStop}%
	\bibitem [{\citenamefont {{Boulanger}}\ \emph {et~al.}(2022)\citenamefont
		{{Boulanger}}, \citenamefont {{Grissonnanche}}, \citenamefont
		{{Lefran{\c{c}}ois}}, \citenamefont {{Gourgout}}, \citenamefont {{Xu}},
		\citenamefont {{Shen}}, \citenamefont {{Greene}},\ and\ \citenamefont
		{{Taillefer}}}]{2022PhRvB.105k5101B}%
	\BibitemOpen
	\bibfield  {author} {\bibinfo {author} {\bibfnamefont {M.-E.}\ \bibnamefont
			{{Boulanger}}}, \bibinfo {author} {\bibfnamefont {G.}~\bibnamefont
			{{Grissonnanche}}}, \bibinfo {author} {\bibfnamefont {{\'E}.}~\bibnamefont
			{{Lefran{\c{c}}ois}}}, \bibinfo {author} {\bibfnamefont {A.}~\bibnamefont
			{{Gourgout}}}, \bibinfo {author} {\bibfnamefont {K.-J.}\ \bibnamefont
			{{Xu}}}, \bibinfo {author} {\bibfnamefont {Z.-X.}\ \bibnamefont {{Shen}}},
		\bibinfo {author} {\bibfnamefont {R.~L.}\ \bibnamefont {{Greene}}},\ and\
		\bibinfo {author} {\bibfnamefont {L.}~\bibnamefont {{Taillefer}}},\
	}\bibfield  {title} {\bibinfo {title} {{Thermal Hall conductivity of
				electron-doped cuprates}},\ }\href
	{https://doi.org/10.1103/PhysRevB.105.115101} {\bibfield  {journal} {\bibinfo
			{journal} {\prb}\ }\textbf {\bibinfo {volume} {105}},\ \bibinfo {eid}
		{115101} (\bibinfo {year} {2022})}\BibitemShut {NoStop}%
	\bibitem [{\citenamefont {{Xu}}\ \emph {et~al.}(2023)\citenamefont {{Xu}},
		\citenamefont {{Carnahan}}, \citenamefont {{Zhang}}, \citenamefont
		{{Sretenovic}}, \citenamefont {{Zhang}}, \citenamefont {{Xiao}},\ and\
		\citenamefont {{Ke}}}]{2023PhRvB.107f0404X}%
	\BibitemOpen
	\bibfield  {author} {\bibinfo {author} {\bibfnamefont {C.}~\bibnamefont
			{{Xu}}}, \bibinfo {author} {\bibfnamefont {C.}~\bibnamefont {{Carnahan}}},
		\bibinfo {author} {\bibfnamefont {H.}~\bibnamefont {{Zhang}}}, \bibinfo
		{author} {\bibfnamefont {M.}~\bibnamefont {{Sretenovic}}}, \bibinfo {author}
		{\bibfnamefont {P.}~\bibnamefont {{Zhang}}}, \bibinfo {author} {\bibfnamefont
			{D.}~\bibnamefont {{Xiao}}},\ and\ \bibinfo {author} {\bibfnamefont
			{X.}~\bibnamefont {{Ke}}},\ }\bibfield  {title} {\bibinfo {title} {{Thermal
				Hall effect in a van der Waals triangular magnet FeCl$_{2}$}},\ }\href
	{https://doi.org/10.1103/PhysRevB.107.L060404} {\bibfield  {journal}
		{\bibinfo  {journal} {\prb}\ }\textbf {\bibinfo {volume} {107}},\ \bibinfo
		{eid} {L060404} (\bibinfo {year} {2023})}\BibitemShut {NoStop}%
	\bibitem [{\citenamefont {{Lefran{\c{c}}ois}}\ \emph
		{et~al.}(2022)\citenamefont {{Lefran{\c{c}}ois}}, \citenamefont
		{{Grissonnanche}}, \citenamefont {{Baglo}}, \citenamefont {{Lampen-Kelley}},
		\citenamefont {{Yan}}, \citenamefont {{Balz}}, \citenamefont {{Mandrus}},
		\citenamefont {{Nagler}}, \citenamefont {{Kim}}, \citenamefont {{Kim}},
		\citenamefont {{Doiron-Leyraud}},\ and\ \citenamefont
		{{Taillefer}}}]{2022PhRvX..12b1025L}%
	\BibitemOpen
	\bibfield  {author} {\bibinfo {author} {\bibfnamefont {{\'E}.}~\bibnamefont
			{{Lefran{\c{c}}ois}}}, \bibinfo {author} {\bibfnamefont {G.}~\bibnamefont
			{{Grissonnanche}}}, \bibinfo {author} {\bibfnamefont {J.}~\bibnamefont
			{{Baglo}}}, \bibinfo {author} {\bibfnamefont {P.}~\bibnamefont
			{{Lampen-Kelley}}}, \bibinfo {author} {\bibfnamefont {J.~Q.}\ \bibnamefont
			{{Yan}}}, \bibinfo {author} {\bibfnamefont {C.}~\bibnamefont {{Balz}}},
		\bibinfo {author} {\bibfnamefont {D.}~\bibnamefont {{Mandrus}}}, \bibinfo
		{author} {\bibfnamefont {S.~E.}\ \bibnamefont {{Nagler}}}, \bibinfo {author}
		{\bibfnamefont {S.}~\bibnamefont {{Kim}}}, \bibinfo {author} {\bibfnamefont
			{Y.-J.}\ \bibnamefont {{Kim}}}, \bibinfo {author} {\bibfnamefont
			{N.}~\bibnamefont {{Doiron-Leyraud}}},\ and\ \bibinfo {author} {\bibfnamefont
			{L.}~\bibnamefont {{Taillefer}}},\ }\bibfield  {title} {\bibinfo {title}
		{{Evidence of a Phonon Hall Effect in the Kitaev Spin Liquid Candidate
				{\ensuremath{\alpha}} -RuCl$_{3}$}},\ }\href
	{https://doi.org/10.1103/PhysRevX.12.021025} {\bibfield  {journal} {\bibinfo
			{journal} {Physical Review X}\ }\textbf {\bibinfo {volume} {12}},\ \bibinfo
		{eid} {021025} (\bibinfo {year} {2022})}\BibitemShut {NoStop}%
	\bibitem [{\citenamefont {{Jiang}}\ \emph {et~al.}(2022)\citenamefont
		{{Jiang}}, \citenamefont {{Li}}, \citenamefont {{Fauqu{\'e}}},\ and\
		\citenamefont {{Behnia}}}]{2022PNAS..11901975J}%
	\BibitemOpen
	\bibfield  {author} {\bibinfo {author} {\bibfnamefont {S.}~\bibnamefont
			{{Jiang}}}, \bibinfo {author} {\bibfnamefont {X.}~\bibnamefont {{Li}}},
		\bibinfo {author} {\bibfnamefont {B.}~\bibnamefont {{Fauqu{\'e}}}},\ and\
		\bibinfo {author} {\bibfnamefont {K.}~\bibnamefont {{Behnia}}},\ }\bibfield
	{title} {\bibinfo {title} {{Phonon drag thermal Hall effect in metallic
				strontium titanate}},\ }\href {https://doi.org/10.1073/pnas.2201975119}
	{\bibfield  {journal} {\bibinfo  {journal} {Proceedings of the National
				Academy of Science}\ }\textbf {\bibinfo {volume} {119}},\ \bibinfo {eid}
		{e2201975119} (\bibinfo {year} {2022})}\BibitemShut {NoStop}%
	\bibitem [{\citenamefont {{Meng}}\ \emph {et~al.}(2024)\citenamefont {{Meng}},
		\citenamefont {{Li}}, \citenamefont {{Zhao}}, \citenamefont {{Dong}},
		\citenamefont {{Zhu}},\ and\ \citenamefont {{Behnia}}}]{2024arXiv240313306M}%
	\BibitemOpen
	\bibfield  {author} {\bibinfo {author} {\bibfnamefont {Q.}~\bibnamefont
			{{Meng}}}, \bibinfo {author} {\bibfnamefont {X.}~\bibnamefont {{Li}}},
		\bibinfo {author} {\bibfnamefont {L.}~\bibnamefont {{Zhao}}}, \bibinfo
		{author} {\bibfnamefont {C.}~\bibnamefont {{Dong}}}, \bibinfo {author}
		{\bibfnamefont {Z.}~\bibnamefont {{Zhu}}},\ and\ \bibinfo {author}
		{\bibfnamefont {K.}~\bibnamefont {{Behnia}}},\ }\bibfield  {title} {\bibinfo
		{title} {{Thermal Hall effect driven by phonon-magnon hybridization in a
				honeycomb antiferromagnet}},\ }\href
	{https://doi.org/10.48550/arXiv.2403.13306} {\bibfield  {journal} {\bibinfo
			{journal} {arXiv e-prints}\ ,\ \bibinfo {eid} {arXiv:2403.13306}} (\bibinfo
		{year} {2024})}\BibitemShut {NoStop}%
	\bibitem [{\citenamefont {{Gillig}}\ \emph {et~al.}(2023)\citenamefont
		{{Gillig}}, \citenamefont {{Hong}}, \citenamefont {{Wellm}}, \citenamefont
		{{Kataev}}, \citenamefont {{Yao}}, \citenamefont {{Li}}, \citenamefont
		{{B{\"u}chner}},\ and\ \citenamefont {{Hess}}}]{2023PhRvR...5d3110G}%
	\BibitemOpen
	\bibfield  {author} {\bibinfo {author} {\bibfnamefont {M.}~\bibnamefont
			{{Gillig}}}, \bibinfo {author} {\bibfnamefont {X.}~\bibnamefont {{Hong}}},
		\bibinfo {author} {\bibfnamefont {C.}~\bibnamefont {{Wellm}}}, \bibinfo
		{author} {\bibfnamefont {V.}~\bibnamefont {{Kataev}}}, \bibinfo {author}
		{\bibfnamefont {W.}~\bibnamefont {{Yao}}}, \bibinfo {author} {\bibfnamefont
			{Y.}~\bibnamefont {{Li}}}, \bibinfo {author} {\bibfnamefont {B.}~\bibnamefont
			{{B{\"u}chner}}},\ and\ \bibinfo {author} {\bibfnamefont {C.}~\bibnamefont
			{{Hess}}},\ }\bibfield  {title} {\bibinfo {title} {{Phononic-magnetic
				dichotomy of the thermal Hall effect in the Kitaev material
				Na$_{2}$Co$_{2}$TeO$_{6}$}},\ }\href
	{https://doi.org/10.1103/PhysRevResearch.5.043110} {\bibfield  {journal}
		{\bibinfo  {journal} {Physical Review Research}\ }\textbf {\bibinfo {volume}
			{5}},\ \bibinfo {eid} {043110} (\bibinfo {year} {2023})}\BibitemShut
	{NoStop}%
	\bibitem [{\citenamefont {{Li}}\ \emph
		{et~al.}(2023{\natexlab{a}})\citenamefont {{Li}}, \citenamefont {{Neumann}},
		\citenamefont {{Guang}}, \citenamefont {{Huang}}, \citenamefont {{Liu}},
		\citenamefont {{Xia}}, \citenamefont {{Yue}}, \citenamefont {{Sun}},
		\citenamefont {{Wang}}, \citenamefont {{Li}}, \citenamefont {{Jiang}},
		\citenamefont {{Fang}}, \citenamefont {{Jiang}}, \citenamefont {{Zhao}},
		\citenamefont {{Mook}}, \citenamefont {{Henk}}, \citenamefont {{Mertig}},
		\citenamefont {{Zhou}},\ and\ \citenamefont {{Sun}}}]{2023PhRvB.108n0402L}%
	\BibitemOpen
	\bibfield  {author} {\bibinfo {author} {\bibfnamefont {N.}~\bibnamefont
			{{Li}}}, \bibinfo {author} {\bibfnamefont {R.~R.}\ \bibnamefont {{Neumann}}},
		\bibinfo {author} {\bibfnamefont {S.~K.}\ \bibnamefont {{Guang}}}, \bibinfo
		{author} {\bibfnamefont {Q.}~\bibnamefont {{Huang}}}, \bibinfo {author}
		{\bibfnamefont {J.}~\bibnamefont {{Liu}}}, \bibinfo {author} {\bibfnamefont
			{K.}~\bibnamefont {{Xia}}}, \bibinfo {author} {\bibfnamefont {X.~Y.}\
			\bibnamefont {{Yue}}}, \bibinfo {author} {\bibfnamefont {Y.}~\bibnamefont
			{{Sun}}}, \bibinfo {author} {\bibfnamefont {Y.~Y.}\ \bibnamefont {{Wang}}},
		\bibinfo {author} {\bibfnamefont {Q.~J.}\ \bibnamefont {{Li}}}, \bibinfo
		{author} {\bibfnamefont {Y.}~\bibnamefont {{Jiang}}}, \bibinfo {author}
		{\bibfnamefont {J.}~\bibnamefont {{Fang}}}, \bibinfo {author} {\bibfnamefont
			{Z.}~\bibnamefont {{Jiang}}}, \bibinfo {author} {\bibfnamefont
			{X.}~\bibnamefont {{Zhao}}}, \bibinfo {author} {\bibfnamefont
			{A.}~\bibnamefont {{Mook}}}, \bibinfo {author} {\bibfnamefont
			{J.}~\bibnamefont {{Henk}}}, \bibinfo {author} {\bibfnamefont
			{I.}~\bibnamefont {{Mertig}}}, \bibinfo {author} {\bibfnamefont {H.~D.}\
			\bibnamefont {{Zhou}}},\ and\ \bibinfo {author} {\bibfnamefont {X.~F.}\
			\bibnamefont {{Sun}}},\ }\bibfield  {title} {\bibinfo {title}
		{{Magnon-polaron driven thermal Hall effect in a Heisenberg-Kitaev
				antiferromagnet}},\ }\href {https://doi.org/10.1103/PhysRevB.108.L140402}
	{\bibfield  {journal} {\bibinfo  {journal} {\prb}\ }\textbf {\bibinfo
			{volume} {108}},\ \bibinfo {eid} {L140402} (\bibinfo {year}
		{2023}{\natexlab{a}})}\BibitemShut {NoStop}%
	\bibitem [{\citenamefont {{Onose}}\ \emph {et~al.}(2010)\citenamefont
		{{Onose}}, \citenamefont {{Ideue}}, \citenamefont {{Katsura}}, \citenamefont
		{{Shiomi}}, \citenamefont {{Nagaosa}},\ and\ \citenamefont
		{{Tokura}}}]{2010Sci...329..297O}%
	\BibitemOpen
	\bibfield  {author} {\bibinfo {author} {\bibfnamefont {Y.}~\bibnamefont
			{{Onose}}}, \bibinfo {author} {\bibfnamefont {T.}~\bibnamefont {{Ideue}}},
		\bibinfo {author} {\bibfnamefont {H.}~\bibnamefont {{Katsura}}}, \bibinfo
		{author} {\bibfnamefont {Y.}~\bibnamefont {{Shiomi}}}, \bibinfo {author}
		{\bibfnamefont {N.}~\bibnamefont {{Nagaosa}}},\ and\ \bibinfo {author}
		{\bibfnamefont {Y.}~\bibnamefont {{Tokura}}},\ }\bibfield  {title} {\bibinfo
		{title} {{Observation of the Magnon Hall Effect}},\ }\href
	{https://doi.org/10.1126/science.1188260} {\bibfield  {journal} {\bibinfo
			{journal} {Science}\ }\textbf {\bibinfo {volume} {329}},\ \bibinfo {pages}
		{297} (\bibinfo {year} {2010})}\BibitemShut {NoStop}%
	\bibitem [{\citenamefont {Ataei}\ \emph {et~al.}(2024)\citenamefont {Ataei},
		\citenamefont {Grissonnanche}, \citenamefont {Boulanger}, \citenamefont
		{Chen}, \citenamefont {LefranÃ§ois}, \citenamefont {Brouet},\ and\
		\citenamefont {Taillefer}}]{Ataei_2024}%
	\BibitemOpen
	\bibfield  {author} {\bibinfo {author} {\bibfnamefont {A.}~\bibnamefont
			{Ataei}}, \bibinfo {author} {\bibfnamefont {G.}~\bibnamefont
			{Grissonnanche}}, \bibinfo {author} {\bibfnamefont {M.-E.}\ \bibnamefont
			{Boulanger}}, \bibinfo {author} {\bibfnamefont {L.}~\bibnamefont {Chen}},
		\bibinfo {author} {\bibfnamefont {Ã.}~\bibnamefont {LefranÃ§ois}}, \bibinfo
		{author} {\bibfnamefont {V.}~\bibnamefont {Brouet}},\ and\ \bibinfo {author}
		{\bibfnamefont {L.}~\bibnamefont {Taillefer}},\ }\bibfield  {title} {\bibinfo
		{title} {{Phonon chirality from impurity scattering in the antiferromagnetic
				phase of Sr$_{2}$IrO$_{4}$}},\ }\href
	{https://doi.org/10.1038/s41567-024-02384-5} {\bibfield  {journal} {\bibinfo
			{journal} {Nature Physics}\ }\textbf {\bibinfo {volume} {20}},\ \bibinfo
		{pages} {585-588} (\bibinfo {year} {2024})}\BibitemShut {NoStop}%
	\bibitem [{\citenamefont {{Akazawa}}\ \emph {et~al.}(2020)\citenamefont
		{{Akazawa}}, \citenamefont {{Shimozawa}}, \citenamefont {{Kittaka}},
		\citenamefont {{Sakakibara}}, \citenamefont {{Okuma}}, \citenamefont
		{{Hiroi}}, \citenamefont {{Lee}}, \citenamefont {{Kawashima}}, \citenamefont
		{{Han}},\ and\ \citenamefont {{Yamashita}}}]{2020PhRvX..10d1059A}%
	\BibitemOpen
	\bibfield  {author} {\bibinfo {author} {\bibfnamefont {M.}~\bibnamefont
			{{Akazawa}}}, \bibinfo {author} {\bibfnamefont {M.}~\bibnamefont
			{{Shimozawa}}}, \bibinfo {author} {\bibfnamefont {S.}~\bibnamefont
			{{Kittaka}}}, \bibinfo {author} {\bibfnamefont {T.}~\bibnamefont
			{{Sakakibara}}}, \bibinfo {author} {\bibfnamefont {R.}~\bibnamefont
			{{Okuma}}}, \bibinfo {author} {\bibfnamefont {Z.}~\bibnamefont {{Hiroi}}},
		\bibinfo {author} {\bibfnamefont {H.-Y.}\ \bibnamefont {{Lee}}}, \bibinfo
		{author} {\bibfnamefont {N.}~\bibnamefont {{Kawashima}}}, \bibinfo {author}
		{\bibfnamefont {J.~H.}\ \bibnamefont {{Han}}},\ and\ \bibinfo {author}
		{\bibfnamefont {M.}~\bibnamefont {{Yamashita}}},\ }\bibfield  {title}
	{\bibinfo {title} {{Thermal Hall Effects of Spins and Phonons in Kagome
				Antiferromagnet Cd-Kapellasite}},\ }\href
	{https://doi.org/10.1103/PhysRevX.10.041059} {\bibfield  {journal} {\bibinfo
			{journal} {Physical Review X}\ }\textbf {\bibinfo {volume} {10}},\ \bibinfo
		{eid} {041059} (\bibinfo {year} {2020})}\BibitemShut {NoStop}%
	\bibitem [{\citenamefont {{Li}}\ \emph {et~al.}(2020)\citenamefont {{Li}},
		\citenamefont {{Fauqu{\'e}}}, \citenamefont {{Zhu}},\ and\ \citenamefont
		{{Behnia}}}]{2020PhRvL.124j5901L}%
	\BibitemOpen
	\bibfield  {author} {\bibinfo {author} {\bibfnamefont {X.}~\bibnamefont
			{{Li}}}, \bibinfo {author} {\bibfnamefont {B.}~\bibnamefont {{Fauqu{\'e}}}},
		\bibinfo {author} {\bibfnamefont {Z.}~\bibnamefont {{Zhu}}},\ and\ \bibinfo
		{author} {\bibfnamefont {K.}~\bibnamefont {{Behnia}}},\ }\bibfield  {title}
	{\bibinfo {title} {{Phonon Thermal Hall Effect in Strontium Titanate}},\
	}\href {https://doi.org/10.1103/PhysRevLett.124.105901} {\bibfield  {journal}
		{\bibinfo  {journal} {\prl}\ }\textbf {\bibinfo {volume} {124}},\ \bibinfo
		{eid} {105901} (\bibinfo {year} {2020})}\BibitemShut {NoStop}%
	\bibitem [{\citenamefont {{Li}}\ \emph
		{et~al.}(2023{\natexlab{b}})\citenamefont {{Li}}, \citenamefont {{Machida}},
		\citenamefont {{Subedi}}, \citenamefont {{Zhu}}, \citenamefont {{Li}},\ and\
		\citenamefont {{Behnia}}}]{2023NatCo..14.1027L}%
	\BibitemOpen
	\bibfield  {author} {\bibinfo {author} {\bibfnamefont {X.}~\bibnamefont
			{{Li}}}, \bibinfo {author} {\bibfnamefont {Y.}~\bibnamefont {{Machida}}},
		\bibinfo {author} {\bibfnamefont {A.}~\bibnamefont {{Subedi}}}, \bibinfo
		{author} {\bibfnamefont {Z.}~\bibnamefont {{Zhu}}}, \bibinfo {author}
		{\bibfnamefont {L.}~\bibnamefont {{Li}}},\ and\ \bibinfo {author}
		{\bibfnamefont {K.}~\bibnamefont {{Behnia}}},\ }\bibfield  {title} {\bibinfo
		{title} {{The phonon thermal Hall angle in black phosphorus}},\ }\href
	{https://doi.org/10.1038/s41467-023-36750-3} {\bibfield  {journal} {\bibinfo
			{journal} {Nature Communications}\ }\textbf {\bibinfo {volume} {14}},\
		\bibinfo {eid} {1027} (\bibinfo {year} {2023}{\natexlab{b}})}\BibitemShut
	{NoStop}%
	\bibitem [{\citenamefont {{Sharma}}\ \emph {et~al.}(2024)\citenamefont
		{{Sharma}}, \citenamefont {{Bagchi}}, \citenamefont {{Wang}}, \citenamefont
		{{Ando}},\ and\ \citenamefont {{Lorenz}}}]{2024PhRvB.109j4304S}%
	\BibitemOpen
	\bibfield  {author} {\bibinfo {author} {\bibfnamefont {R.}~\bibnamefont
			{{Sharma}}}, \bibinfo {author} {\bibfnamefont {M.}~\bibnamefont {{Bagchi}}},
		\bibinfo {author} {\bibfnamefont {Y.}~\bibnamefont {{Wang}}}, \bibinfo
		{author} {\bibfnamefont {Y.}~\bibnamefont {{Ando}}},\ and\ \bibinfo {author}
		{\bibfnamefont {T.}~\bibnamefont {{Lorenz}}},\ }\bibfield  {title} {\bibinfo
		{title} {{Phonon thermal Hall effect in charge-compensated topological
				insulators}},\ }\href {https://doi.org/10.1103/PhysRevB.109.104304}
	{\bibfield  {journal} {\bibinfo  {journal} {\prb}\ }\textbf {\bibinfo
			{volume} {109}},\ \bibinfo {eid} {104304} (\bibinfo {year}
		{2024})}\BibitemShut {NoStop}%
	\bibitem [{\citenamefont {{Jin}}\ \emph {et~al.}(2024)\citenamefont {{Jin}},
		\citenamefont {{Zhang}}, \citenamefont {{Wan}}, \citenamefont {{Wang}},
		\citenamefont {{Jiao}},\ and\ \citenamefont {{Li}}}]{2024arXiv240402863J}%
	\BibitemOpen
	\bibfield  {author} {\bibinfo {author} {\bibfnamefont {X.}~\bibnamefont
			{{Jin}}}, \bibinfo {author} {\bibfnamefont {X.}~\bibnamefont {{Zhang}}},
		\bibinfo {author} {\bibfnamefont {W.}~\bibnamefont {{Wan}}}, \bibinfo
		{author} {\bibfnamefont {H.}~\bibnamefont {{Wang}}}, \bibinfo {author}
		{\bibfnamefont {Y.}~\bibnamefont {{Jiao}}},\ and\ \bibinfo {author}
		{\bibfnamefont {S.}~\bibnamefont {{Li}}},\ }\bibfield  {title} {\bibinfo
		{title} {{Discovery of universal phonon thermal Hall effect in crystals}},\
	}\href {https://doi.org/10.48550/arXiv.2404.02863} {\bibfield  {journal}
		{\bibinfo  {journal} {arXiv e-prints}\ ,\ \bibinfo {eid} {arXiv:2404.02863}}
		(\bibinfo {year} {2024})}\BibitemShut {NoStop}%
	\bibitem [{\citenamefont {{Castelnovo}}\ \emph {et~al.}(2008)\citenamefont
		{{Castelnovo}}, \citenamefont {{Moessner}},\ and\ \citenamefont
		{{Sondhi}}}]{2008Natur.451...42C}%
	\BibitemOpen
	\bibfield  {author} {\bibinfo {author} {\bibfnamefont {C.}~\bibnamefont
			{{Castelnovo}}}, \bibinfo {author} {\bibfnamefont {R.}~\bibnamefont
			{{Moessner}}},\ and\ \bibinfo {author} {\bibfnamefont {S.~L.}\ \bibnamefont
			{{Sondhi}}},\ }\bibfield  {title} {\bibinfo {title} {{Magnetic monopoles in
				spin ice}},\ }\href {https://doi.org/10.1038/nature06433} {\bibfield
		{journal} {\bibinfo  {journal} {\nat}\ }\textbf {\bibinfo {volume} {451}},\
		\bibinfo {pages} {42} (\bibinfo {year} {2008})}\BibitemShut {NoStop}%
	\bibitem [{\citenamefont {{Morris}}\ \emph {et~al.}(2009)\citenamefont
		{{Morris}}, \citenamefont {{Tennant}}, \citenamefont {{Grigera}},
		\citenamefont {{Klemke}}, \citenamefont {{Castelnovo}}, \citenamefont
		{{Moessner}}, \citenamefont {{Czternasty}}, \citenamefont {{Meissner}},
		\citenamefont {{Rule}}, \citenamefont {{Hoffmann}}, \citenamefont {{Kiefer}},
		\citenamefont {{Gerischer}}, \citenamefont {{Slobinsky}},\ and\ \citenamefont
		{{Perry}}}]{2009Sci...326..411M}%
	\BibitemOpen
	\bibfield  {author} {\bibinfo {author} {\bibfnamefont {D.~J.~P.}\
			\bibnamefont {{Morris}}}, \bibinfo {author} {\bibfnamefont {D.~A.}\
			\bibnamefont {{Tennant}}}, \bibinfo {author} {\bibfnamefont {S.~A.}\
			\bibnamefont {{Grigera}}}, \bibinfo {author} {\bibfnamefont {B.}~\bibnamefont
			{{Klemke}}}, \bibinfo {author} {\bibfnamefont {C.}~\bibnamefont
			{{Castelnovo}}}, \bibinfo {author} {\bibfnamefont {R.}~\bibnamefont
			{{Moessner}}}, \bibinfo {author} {\bibfnamefont {C.}~\bibnamefont
			{{Czternasty}}}, \bibinfo {author} {\bibfnamefont {M.}~\bibnamefont
			{{Meissner}}}, \bibinfo {author} {\bibfnamefont {K.~C.}\ \bibnamefont
			{{Rule}}}, \bibinfo {author} {\bibfnamefont {J.~U.}\ \bibnamefont
			{{Hoffmann}}}, \bibinfo {author} {\bibfnamefont {K.}~\bibnamefont
			{{Kiefer}}}, \bibinfo {author} {\bibfnamefont {S.}~\bibnamefont
			{{Gerischer}}}, \bibinfo {author} {\bibfnamefont {D.}~\bibnamefont
			{{Slobinsky}}},\ and\ \bibinfo {author} {\bibfnamefont {R.~S.}\ \bibnamefont
			{{Perry}}},\ }\bibfield  {title} {\bibinfo {title} {{Dirac Strings and
				Magnetic Monopoles in the Spin Ice Dy$_{2}$Ti$_{2}$O$_{7}$}},\ }\href
	{https://doi.org/10.1126/science.1178868} {\bibfield  {journal} {\bibinfo
			{journal} {Science}\ }\textbf {\bibinfo {volume} {326}},\ \bibinfo {pages}
		{411} (\bibinfo {year} {2009})}\BibitemShut {NoStop}%
	\bibitem [{\citenamefont {{Paulsen}}\ \emph {et~al.}(2014)\citenamefont
		{{Paulsen}}, \citenamefont {{Jackson}}, \citenamefont {{Lhotel}},
		\citenamefont {{Canals}}, \citenamefont {{Prabhakaran}}, \citenamefont
		{{Matsuhira}}, \citenamefont {{Giblin}},\ and\ \citenamefont
		{{Bramwell}}}]{2014NatPh..10..135P}%
	\BibitemOpen
	\bibfield  {author} {\bibinfo {author} {\bibfnamefont {C.}~\bibnamefont
			{{Paulsen}}}, \bibinfo {author} {\bibfnamefont {M.~J.}\ \bibnamefont
			{{Jackson}}}, \bibinfo {author} {\bibfnamefont {E.}~\bibnamefont {{Lhotel}}},
		\bibinfo {author} {\bibfnamefont {B.}~\bibnamefont {{Canals}}}, \bibinfo
		{author} {\bibfnamefont {D.}~\bibnamefont {{Prabhakaran}}}, \bibinfo {author}
		{\bibfnamefont {K.}~\bibnamefont {{Matsuhira}}}, \bibinfo {author}
		{\bibfnamefont {S.~R.}\ \bibnamefont {{Giblin}}},\ and\ \bibinfo {author}
		{\bibfnamefont {S.~T.}\ \bibnamefont {{Bramwell}}},\ }\bibfield  {title}
	{\bibinfo {title} {{Far-from-equilibrium monopole dynamics in spin ice}},\
	}\href {https://doi.org/10.1038/nphys2847} {\bibfield  {journal} {\bibinfo
			{journal} {Nature Physics}\ }\textbf {\bibinfo {volume} {10}},\ \bibinfo
		{pages} {135} (\bibinfo {year} {2014})}\BibitemShut {NoStop}%
	\bibitem [{\citenamefont {{Bramwell}}\ and\ \citenamefont
		{{Gingras}}(2001)}]{2001Sci...294.1495B}%
	\BibitemOpen
	\bibfield  {author} {\bibinfo {author} {\bibfnamefont {S.~T.}\ \bibnamefont
			{{Bramwell}}}\ and\ \bibinfo {author} {\bibfnamefont {M.~J.~P.}\ \bibnamefont
			{{Gingras}}},\ }\bibfield  {title} {\bibinfo {title} {{Spin Ice State in
				Frustrated Magnetic Pyrochlore Materials}},\ }\href
	{https://doi.org/10.1126/science.1064761} {\bibfield  {journal} {\bibinfo
			{journal} {Science}\ }\textbf {\bibinfo {volume} {294}},\ \bibinfo {pages}
		{1495} (\bibinfo {year} {2001})}\BibitemShut {NoStop}%
	\bibitem [{\citenamefont {Pomaranski}\ \emph {et~al.}(2013)\citenamefont
		{Pomaranski}, \citenamefont {Yaraskavitch}, \citenamefont {Meng},
		\citenamefont {Ross}, \citenamefont {Noad}, \citenamefont {Dabkowska},
		\citenamefont {Gaulin},\ and\ \citenamefont {Kycia}}]{pomaranski2013absence}%
	\BibitemOpen
	\bibfield  {author} {\bibinfo {author} {\bibfnamefont {D.}~\bibnamefont
			{Pomaranski}}, \bibinfo {author} {\bibfnamefont {L.}~\bibnamefont
			{Yaraskavitch}}, \bibinfo {author} {\bibfnamefont {S.}~\bibnamefont {Meng}},
		\bibinfo {author} {\bibfnamefont {K.}~\bibnamefont {Ross}}, \bibinfo {author}
		{\bibfnamefont {H.}~\bibnamefont {Noad}}, \bibinfo {author} {\bibfnamefont
			{H.}~\bibnamefont {Dabkowska}}, \bibinfo {author} {\bibfnamefont
			{B.}~\bibnamefont {Gaulin}},\ and\ \bibinfo {author} {\bibfnamefont
			{J.}~\bibnamefont {Kycia}},\ }\bibfield  {title} {\bibinfo {title} {{Absence
				of Pauling's residual entropy in thermally equilibrated
				Dy$_{2}$Ti$_{2}$O$_{7}$}},\ }\href {https://doi.org/10.1038/nphys2591}
	{\bibfield  {journal} {\bibinfo  {journal} {Nature Physics}\ }\textbf
		{\bibinfo {volume} {9}},\ \bibinfo {pages} {353} (\bibinfo {year}
		{2013})}\BibitemShut {NoStop}%
	\bibitem [{\citenamefont {Ramirez}\ \emph {et~al.}(1999)\citenamefont
		{Ramirez}, \citenamefont {Hayashi}, \citenamefont {Cava}, \citenamefont
		{Siddharthan},\ and\ \citenamefont {Shastry}}]{ramirez1999zero}%
	\BibitemOpen
	\bibfield  {author} {\bibinfo {author} {\bibfnamefont {A.~P.}\ \bibnamefont
			{Ramirez}}, \bibinfo {author} {\bibfnamefont {A.}~\bibnamefont {Hayashi}},
		\bibinfo {author} {\bibfnamefont {R.~J.}\ \bibnamefont {Cava}}, \bibinfo
		{author} {\bibfnamefont {R.}~\bibnamefont {Siddharthan}},\ and\ \bibinfo
		{author} {\bibfnamefont {B.}~\bibnamefont {Shastry}},\ }\bibfield  {title}
	{\bibinfo {title} {Zero-point entropy in spin ice},\ }\href
	{https://doi.org/10.1038/20619} {\bibfield  {journal} {\bibinfo  {journal}
			{Nature}\ }\textbf {\bibinfo {volume} {399}},\ \bibinfo {pages} {333}
		(\bibinfo {year} {1999})}\BibitemShut {NoStop}%
	\bibitem [{\citenamefont {{Kolland}}\ \emph {et~al.}(2013)\citenamefont
		{{Kolland}}, \citenamefont {{Valldor}}, \citenamefont {{Hiertz}},
		\citenamefont {{Frielingsdorf}},\ and\ \citenamefont
		{{Lorenz}}}]{2013PhRvB..88e4406K}%
	\BibitemOpen
	\bibfield  {author} {\bibinfo {author} {\bibfnamefont {G.}~\bibnamefont
			{{Kolland}}}, \bibinfo {author} {\bibfnamefont {M.}~\bibnamefont
			{{Valldor}}}, \bibinfo {author} {\bibfnamefont {M.}~\bibnamefont {{Hiertz}}},
		\bibinfo {author} {\bibfnamefont {J.}~\bibnamefont {{Frielingsdorf}}},\ and\
		\bibinfo {author} {\bibfnamefont {T.}~\bibnamefont {{Lorenz}}},\ }\bibfield
	{title} {\bibinfo {title} {{Anisotropic heat transport via monopoles in the
				spin-ice compound Dy$_{2}$Ti$_{2}$O$_{7}$}},\ }\href
	{https://doi.org/10.1103/PhysRevB.88.054406} {\bibfield  {journal} {\bibinfo
			{journal} {\prb}\ }\textbf {\bibinfo {volume} {88}},\ \bibinfo {eid} {054406}
		(\bibinfo {year} {2013})}\BibitemShut {NoStop}%
	\bibitem [{\citenamefont {{Kolland}}\ \emph {et~al.}(2012)\citenamefont
		{{Kolland}}, \citenamefont {{Breunig}}, \citenamefont {{Valldor}},
		\citenamefont {{Hiertz}}, \citenamefont {{Frielingsdorf}},\ and\
		\citenamefont {{Lorenz}}}]{2012PhRvB..86f0402K}%
	\BibitemOpen
	\bibfield  {author} {\bibinfo {author} {\bibfnamefont {G.}~\bibnamefont
			{{Kolland}}}, \bibinfo {author} {\bibfnamefont {O.}~\bibnamefont
			{{Breunig}}}, \bibinfo {author} {\bibfnamefont {M.}~\bibnamefont
			{{Valldor}}}, \bibinfo {author} {\bibfnamefont {M.}~\bibnamefont {{Hiertz}}},
		\bibinfo {author} {\bibfnamefont {J.}~\bibnamefont {{Frielingsdorf}}},\ and\
		\bibinfo {author} {\bibfnamefont {T.}~\bibnamefont {{Lorenz}}},\ }\bibfield
	{title} {\bibinfo {title} {{Thermal conductivity and specific heat of the
				spin-ice compound Dy$_{2}$Ti$_{2}$O$_{7}$: Experimental evidence for monopole
				heat transport}},\ }\href {https://doi.org/10.1103/PhysRevB.86.060402}
	{\bibfield  {journal} {\bibinfo  {journal} {\prb}\ }\textbf {\bibinfo
			{volume} {86}},\ \bibinfo {eid} {060402(R)} (\bibinfo {year}
		{2012})}\BibitemShut {NoStop}%
	\bibitem [{\citenamefont {Scharffe}\ \emph {et~al.}(2015)\citenamefont
		{Scharffe}, \citenamefont {Kolland}, \citenamefont {Valldor}, \citenamefont
		{Cho}, \citenamefont {Welter},\ and\ \citenamefont
		{Lorenz}}]{scharffe2015heat}%
	\BibitemOpen
	\bibfield  {author} {\bibinfo {author} {\bibfnamefont {S.}~\bibnamefont
			{Scharffe}}, \bibinfo {author} {\bibfnamefont {G.}~\bibnamefont {Kolland}},
		\bibinfo {author} {\bibfnamefont {M.}~\bibnamefont {Valldor}}, \bibinfo
		{author} {\bibfnamefont {V.}~\bibnamefont {Cho}}, \bibinfo {author}
		{\bibfnamefont {J.}~\bibnamefont {Welter}},\ and\ \bibinfo {author}
		{\bibfnamefont {T.}~\bibnamefont {Lorenz}},\ }\bibfield  {title} {\bibinfo
		{title} {{Heat transport of the spin-ice materials Ho$_{2}$Ti$_{2}$O$_{7}$
				and Dy$_{2}$Ti$_{2}$O$_{7}$}},\ }\href
	{https://doi.org/10.1016/j.jmmm.2014.11.015} {\bibfield  {journal} {\bibinfo
			{journal} {Journal of Magnetism and Magnetic Materials}\ }\textbf {\bibinfo
			{volume} {383}},\ \bibinfo {pages} {83} (\bibinfo {year} {2015})}\BibitemShut
	{NoStop}%
	\bibitem [{\citenamefont {{Scharffe}}\ \emph {et~al.}(2015)\citenamefont
		{{Scharffe}}, \citenamefont {{Breunig}}, \citenamefont {{Cho}}, \citenamefont
		{{Laschitzky}}, \citenamefont {{Valldor}}, \citenamefont {{Welter}},\ and\
		\citenamefont {{Lorenz}}}]{2015PhRvB..92r0405S}%
	\BibitemOpen
	\bibfield  {author} {\bibinfo {author} {\bibfnamefont {S.}~\bibnamefont
			{{Scharffe}}}, \bibinfo {author} {\bibfnamefont {O.}~\bibnamefont
			{{Breunig}}}, \bibinfo {author} {\bibfnamefont {V.}~\bibnamefont {{Cho}}},
		\bibinfo {author} {\bibfnamefont {P.}~\bibnamefont {{Laschitzky}}}, \bibinfo
		{author} {\bibfnamefont {M.}~\bibnamefont {{Valldor}}}, \bibinfo {author}
		{\bibfnamefont {J.~F.}\ \bibnamefont {{Welter}}},\ and\ \bibinfo {author}
		{\bibfnamefont {T.}~\bibnamefont {{Lorenz}}},\ }\bibfield  {title} {\bibinfo
		{title} {{Suppression of Pauling's residual entropy in the dilute spin ice
				(Dy$_{1-x}$Y$_{x}$)$_2$Ti$_{2}$O$_{7}$}},\ }\href
	{https://doi.org/10.1103/PhysRevB.92.180405} {\bibfield  {journal} {\bibinfo
			{journal} {\prb}\ }\textbf {\bibinfo {volume} {92}},\ \bibinfo {eid} {180405(R)}
		(\bibinfo {year} {2015})}\BibitemShut {NoStop}%
	\bibitem [{\citenamefont {{Fukazawa}}\ \emph {et~al.}(2002)\citenamefont
		{{Fukazawa}}, \citenamefont {{Melko}}, \citenamefont {{Higashinaka}},
		\citenamefont {{Maeno}},\ and\ \citenamefont
		{{Gingras}}}]{2002PhRvB..65e4410F}%
	\BibitemOpen
	\bibfield  {author} {\bibinfo {author} {\bibfnamefont {H.}~\bibnamefont
			{{Fukazawa}}}, \bibinfo {author} {\bibfnamefont {R.~G.}\ \bibnamefont
			{{Melko}}}, \bibinfo {author} {\bibfnamefont {R.}~\bibnamefont
			{{Higashinaka}}}, \bibinfo {author} {\bibfnamefont {Y.}~\bibnamefont
			{{Maeno}}},\ and\ \bibinfo {author} {\bibfnamefont {M.~J.~P.}\ \bibnamefont
			{{Gingras}}},\ }\bibfield  {title} {\bibinfo {title} {{Magnetic anisotropy of
				the spin-ice compound Dy$_{2}$Ti$_{2}$O$_{7}$}},\ }\href
	{https://doi.org/10.1103/PhysRevB.65.054410} {\bibfield  {journal} {\bibinfo
			{journal} {\prb}\ }\textbf {\bibinfo {volume} {65}},\ \bibinfo {eid} {054410}
		(\bibinfo {year} {2002})}\BibitemShut {NoStop}%
	\bibitem [{\citenamefont {Toews}\ \emph {et~al.}(2018)\citenamefont {Toews},
		\citenamefont {Reid}, \citenamefont {Nadas}, \citenamefont {Rahemtulla},
		\citenamefont {Kycia}, \citenamefont {Munsie}, \citenamefont {Dabkowska},
		\citenamefont {Gaulin},\ and\ \citenamefont {Hill}}]{PhysRevB.98.134446}%
	\BibitemOpen
	\bibfield  {author} {\bibinfo {author} {\bibfnamefont {W.~H.}\ \bibnamefont
			{Toews}}, \bibinfo {author} {\bibfnamefont {J.~A.}\ \bibnamefont {Reid}},
		\bibinfo {author} {\bibfnamefont {R.~B.}\ \bibnamefont {Nadas}}, \bibinfo
		{author} {\bibfnamefont {A.}~\bibnamefont {Rahemtulla}}, \bibinfo {author}
		{\bibfnamefont {S.}~\bibnamefont {Kycia}}, \bibinfo {author} {\bibfnamefont
			{T.~J.~S.}\ \bibnamefont {Munsie}}, \bibinfo {author} {\bibfnamefont {H.~A.}\
			\bibnamefont {Dabkowska}}, \bibinfo {author} {\bibfnamefont {B.~D.}\
			\bibnamefont {Gaulin}},\ and\ \bibinfo {author} {\bibfnamefont {R.~W.}\
			\bibnamefont {Hill}},\ }\bibfield  {title} {\bibinfo {title} {{Disorder
				dependence of monopole dynamics in
				${\mathrm{Dy}}_{2}{\mathrm{Ti}}_{2}{\mathrm{O}}_{7}$ probed via thermal
				transport measurements}},\ }\href
	{https://doi.org/10.1103/PhysRevB.98.134446} {\bibfield  {journal} {\bibinfo
			{journal} {Phys. Rev. B}\ }\textbf {\bibinfo {volume} {98}},\ \bibinfo
		{pages} {134446} (\bibinfo {year} {2018})}\BibitemShut {NoStop}%
	\bibitem [{\citenamefont {Berman}\ and\ \citenamefont
		{Berman}(1976)}]{berman1976thermal}%
	\BibitemOpen
	\bibfield  {author} {\bibinfo {author} {\bibfnamefont {R.}~\bibnamefont
			{Berman}}\ and\ \bibinfo {author} {\bibfnamefont {R.}~\bibnamefont
			{Berman}},\ }\href {https://books.google.de/books?id=rUh5AAAAIAAJ} {\emph
		{\bibinfo {title} {Thermal Conduction in Solids}}},\ Oxford studies in
	physics\ (\bibinfo  {publisher} {Clarendon Press},\ \bibinfo {year}
	{1976})\BibitemShut {NoStop}%
	\bibitem [{\citenamefont {Kittel}\ and\ \citenamefont
		{McEuen}(2018)}]{kittel2018introduction}%
	\BibitemOpen
	\bibfield  {author} {\bibinfo {author} {\bibfnamefont {C.}~\bibnamefont
			{Kittel}}\ and\ \bibinfo {author} {\bibfnamefont {P.}~\bibnamefont
			{McEuen}},\ }\href@noop {} {\emph {\bibinfo {title} {Introduction to solid
				state physics}}}\ (\bibinfo  {publisher} {John Wiley \& Sons},\ \bibinfo
	{year} {2018})\BibitemShut {NoStop}%
	\bibitem [{\citenamefont {{Li}}\ \emph {et~al.}(2013)\citenamefont {{Li}},
		\citenamefont {{Zhao}}, \citenamefont {{Fan}}, \citenamefont {{Zhang}},
		\citenamefont {{Zhou}}, \citenamefont {{Zhao}},\ and\ \citenamefont
		{{Sun}}}]{2013PhRvB..87u4408L}%
	\BibitemOpen
	\bibfield  {author} {\bibinfo {author} {\bibfnamefont {Q.~J.}\ \bibnamefont
			{{Li}}}, \bibinfo {author} {\bibfnamefont {Z.~Y.}\ \bibnamefont {{Zhao}}},
		\bibinfo {author} {\bibfnamefont {C.}~\bibnamefont {{Fan}}}, \bibinfo
		{author} {\bibfnamefont {F.~B.}\ \bibnamefont {{Zhang}}}, \bibinfo {author}
		{\bibfnamefont {H.~D.}\ \bibnamefont {{Zhou}}}, \bibinfo {author}
		{\bibfnamefont {X.}~\bibnamefont {{Zhao}}},\ and\ \bibinfo {author}
		{\bibfnamefont {X.~F.}\ \bibnamefont {{Sun}}},\ }\bibfield  {title} {\bibinfo
		{title} {{Phonon-glass-like behavior of magnetic origin in single-crystal
				Tb$_{2}$Ti$_{2}$O$_{7}$}},\ }\href
	{https://doi.org/10.1103/PhysRevB.87.214408} {\bibfield  {journal} {\bibinfo
			{journal} {\prb}\ }\textbf {\bibinfo {volume} {87}},\ \bibinfo {eid} {214408}
		(\bibinfo {year} {2013})}\BibitemShut {NoStop}%
	\bibitem [{\citenamefont {{Li}}\ \emph {et~al.}(2015)\citenamefont {{Li}},
		\citenamefont {{Zhao}}, \citenamefont {{Fan}}, \citenamefont {{Tong}},
		\citenamefont {{Zhang}}, \citenamefont {{Shi}}, \citenamefont {{Wu}},
		\citenamefont {{Liu}}, \citenamefont {{Zhou}}, \citenamefont {{Zhao}},\ and\
		\citenamefont {{Sun}}}]{2015PhRvB..92i4408L}%
	\BibitemOpen
	\bibfield  {author} {\bibinfo {author} {\bibfnamefont {S.~J.}\ \bibnamefont
			{{Li}}}, \bibinfo {author} {\bibfnamefont {Z.~Y.}\ \bibnamefont {{Zhao}}},
		\bibinfo {author} {\bibfnamefont {C.}~\bibnamefont {{Fan}}}, \bibinfo
		{author} {\bibfnamefont {B.}~\bibnamefont {{Tong}}}, \bibinfo {author}
		{\bibfnamefont {F.~B.}\ \bibnamefont {{Zhang}}}, \bibinfo {author}
		{\bibfnamefont {J.}~\bibnamefont {{Shi}}}, \bibinfo {author} {\bibfnamefont
			{J.~C.}\ \bibnamefont {{Wu}}}, \bibinfo {author} {\bibfnamefont {X.~G.}\
			\bibnamefont {{Liu}}}, \bibinfo {author} {\bibfnamefont {H.~D.}\ \bibnamefont
			{{Zhou}}}, \bibinfo {author} {\bibfnamefont {X.}~\bibnamefont {{Zhao}}},\
		and\ \bibinfo {author} {\bibfnamefont {X.~F.}\ \bibnamefont {{Sun}}},\
	}\bibfield  {title} {\bibinfo {title} {{Low-temperature thermal conductivity
				of Dy$_{2}$Ti$_{2}$O$_{7}$ and Yb$_{2}$Ti$_{2}$O$_{7}$ single crystals}},\
	}\href {https://doi.org/10.1103/PhysRevB.92.094408} {\bibfield  {journal}
		{\bibinfo  {journal} {\prb}\ }\textbf {\bibinfo {volume} {92}},\ \bibinfo
		{eid} {094408} (\bibinfo {year} {2015})}\BibitemShut {NoStop}%
	\bibitem [{\citenamefont {{Vallipuram}}\ \emph {et~al.}(2023)\citenamefont
		{{Vallipuram}}, \citenamefont {{Chen}}, \citenamefont {{Campillo}},
		\citenamefont {{Mezidi}}, \citenamefont {{Grissonnanche}}, \citenamefont
		{{Zic}}, \citenamefont {{Li}}, \citenamefont {{Fisher}}, \citenamefont
		{{Baglo}},\ and\ \citenamefont {{Taillefer}}}]{2023arXiv231010643V5}%
	\BibitemOpen
	\bibfield  {author} {\bibinfo {author} {\bibfnamefont {A.}~\bibnamefont
			{{Vallipuram}}}, \bibinfo {author} {\bibfnamefont {L.}~\bibnamefont
			{{Chen}}}, \bibinfo {author} {\bibfnamefont {E.}~\bibnamefont {{Campillo}}},
		\bibinfo {author} {\bibfnamefont {M.}~\bibnamefont {{Mezidi}}}, \bibinfo
		{author} {\bibfnamefont {G.}~\bibnamefont {{Grissonnanche}}}, \bibinfo
		{author} {\bibfnamefont {M.~P.}\ \bibnamefont {{Zic}}}, \bibinfo {author}
		{\bibfnamefont {Y.}~\bibnamefont {{Li}}}, \bibinfo {author} {\bibfnamefont
			{I.~R.}\ \bibnamefont {{Fisher}}}, \bibinfo {author} {\bibfnamefont
			{J.}~\bibnamefont {{Baglo}}},\ and\ \bibinfo {author} {\bibfnamefont
			{L.}~\bibnamefont {{Taillefer}}},\ }\bibfield  {title} {\bibinfo {title}
		{{Role of magnetic ions in the thermal Hall effect of the paramagnetic
				insulator TmVO$_{4}$}},\ }\href {https://doi.org/10.48550/arXiv.2310.10643}
	{\bibfield  {journal} {\bibinfo  {journal} {arXiv e-prints}\ ,\ \bibinfo
			{eid} {arXiv:2310.10643}} (\bibinfo {year} {2023})}\BibitemShut {NoStop}%
	\bibitem [{\citenamefont {{Sharma}}\ \emph {et~al.}(2004)\citenamefont
		{{Sharma}}, \citenamefont {{Ahn}}, \citenamefont {{Hur}}, \citenamefont
		{{Park}}, \citenamefont {{Kim}}, \citenamefont {{Lee}}, \citenamefont
		{{Park}}, \citenamefont {{Guha}},\ and\ \citenamefont
		{{Cheong}}}]{2004PhRvL..93q7202S}%
	\BibitemOpen
	\bibfield  {author} {\bibinfo {author} {\bibfnamefont {P.~A.}\ \bibnamefont
			{{Sharma}}}, \bibinfo {author} {\bibfnamefont {J.~S.}\ \bibnamefont {{Ahn}}},
		\bibinfo {author} {\bibfnamefont {N.}~\bibnamefont {{Hur}}}, \bibinfo
		{author} {\bibfnamefont {S.}~\bibnamefont {{Park}}}, \bibinfo {author}
		{\bibfnamefont {S.~B.}\ \bibnamefont {{Kim}}}, \bibinfo {author}
		{\bibfnamefont {S.}~\bibnamefont {{Lee}}}, \bibinfo {author} {\bibfnamefont
			{J.~G.}\ \bibnamefont {{Park}}}, \bibinfo {author} {\bibfnamefont
			{S.}~\bibnamefont {{Guha}}},\ and\ \bibinfo {author} {\bibfnamefont {S.~W.}\
			\bibnamefont {{Cheong}}},\ }\bibfield  {title} {\bibinfo {title} {{Thermal
				Conductivity of Geometrically Frustrated, Ferroelectric YMnO$_{3}$:
				Extraordinary Spin-Phonon Interactions}},\ }\href
	{https://doi.org/10.1103/PhysRevLett.93.177202} {\bibfield  {journal}
		{\bibinfo  {journal} {\prl}\ }\textbf {\bibinfo {volume} {93}},\ \bibinfo
		{eid} {177202} (\bibinfo {year} {2004})}\BibitemShut {NoStop}%
	\bibitem [{\citenamefont {{Berggold}}\ \emph {et~al.}(2007)\citenamefont
		{{Berggold}}, \citenamefont {{Baier}}, \citenamefont {{Meier}}, \citenamefont
		{{Mydosh}}, \citenamefont {{Lorenz}}, \citenamefont {{Hemberger}},
		\citenamefont {{Balbashov}}, \citenamefont {{Aliouane}},\ and\ \citenamefont
		{{Argyriou}}}]{2007PhRvB..76i4418B}%
	\BibitemOpen
	\bibfield  {author} {\bibinfo {author} {\bibfnamefont {K.}~\bibnamefont
			{{Berggold}}}, \bibinfo {author} {\bibfnamefont {J.}~\bibnamefont {{Baier}}},
		\bibinfo {author} {\bibfnamefont {D.}~\bibnamefont {{Meier}}}, \bibinfo
		{author} {\bibfnamefont {J.~A.}\ \bibnamefont {{Mydosh}}}, \bibinfo {author}
		{\bibfnamefont {T.}~\bibnamefont {{Lorenz}}}, \bibinfo {author}
		{\bibfnamefont {J.}~\bibnamefont {{Hemberger}}}, \bibinfo {author}
		{\bibfnamefont {A.}~\bibnamefont {{Balbashov}}}, \bibinfo {author}
		{\bibfnamefont {N.}~\bibnamefont {{Aliouane}}},\ and\ \bibinfo {author}
		{\bibfnamefont {D.~N.}\ \bibnamefont {{Argyriou}}},\ }\bibfield  {title}
	{\bibinfo {title} {{Anomalous thermal expansion and strong damping of the
				thermal conductivity of NdMnO$_{3}$ and TbMnO$_{3}$ due to 4f crystal-field
				excitations}},\ }\href {https://doi.org/10.1103/PhysRevB.76.094418}
	{\bibfield  {journal} {\bibinfo  {journal} {\prb}\ }\textbf {\bibinfo
			{volume} {76}},\ \bibinfo {eid} {094418} (\bibinfo {year}
		{2007})}\BibitemShut {NoStop}%
	\bibitem [{\citenamefont {{Tokiwa}}\ \emph {et~al.}(2016)\citenamefont
		{{Tokiwa}}, \citenamefont {{Yamashita}}, \citenamefont {{Udagawa}},
		\citenamefont {{Kittaka}}, \citenamefont {{Sakakibara}}, \citenamefont
		{{Terazawa}}, \citenamefont {{Shimoyama}}, \citenamefont {{Terashima}},
		\citenamefont {{Yasui}}, \citenamefont {{Shibauchi}},\ and\ \citenamefont
		{{Matsuda}}}]{2016NatCo...710807T}%
	\BibitemOpen
	\bibfield  {author} {\bibinfo {author} {\bibfnamefont {Y.}~\bibnamefont
			{{Tokiwa}}}, \bibinfo {author} {\bibfnamefont {T.}~\bibnamefont
			{{Yamashita}}}, \bibinfo {author} {\bibfnamefont {M.}~\bibnamefont
			{{Udagawa}}}, \bibinfo {author} {\bibfnamefont {S.}~\bibnamefont
			{{Kittaka}}}, \bibinfo {author} {\bibfnamefont {T.}~\bibnamefont
			{{Sakakibara}}}, \bibinfo {author} {\bibfnamefont {D.}~\bibnamefont
			{{Terazawa}}}, \bibinfo {author} {\bibfnamefont {Y.}~\bibnamefont
			{{Shimoyama}}}, \bibinfo {author} {\bibfnamefont {T.}~\bibnamefont
			{{Terashima}}}, \bibinfo {author} {\bibfnamefont {Y.}~\bibnamefont
			{{Yasui}}}, \bibinfo {author} {\bibfnamefont {T.}~\bibnamefont
			{{Shibauchi}}},\ and\ \bibinfo {author} {\bibfnamefont {Y.}~\bibnamefont
			{{Matsuda}}},\ }\bibfield  {title} {\bibinfo {title} {{Possible observation
				of highly itinerant quantum magnetic monopoles in the frustrated pyrochlore
				Yb$_{2}$Ti$_{2}$O$_{7}$}},\ }\href {https://doi.org/10.1038/ncomms10807}
	{\bibfield  {journal} {\bibinfo  {journal} {Nature Communications}\ }\textbf
		{\bibinfo {volume} {7}},\ \bibinfo {eid} {10807} (\bibinfo {year}
		{2016})}\BibitemShut {NoStop}%
	\bibitem [{\citenamefont {{Ruminy}}\ \emph {et~al.}(2016)\citenamefont
		{{Ruminy}}, \citenamefont {{Valdez}}, \citenamefont {{Wehinger}},
		\citenamefont {{Bosak}}, \citenamefont {{Adroja}}, \citenamefont {{Stuhr}},
		\citenamefont {{Iida}}, \citenamefont {{Kamazawa}}, \citenamefont
		{{Pomjakushina}}, \citenamefont {{Prabakharan}}, \citenamefont {{Haas}},
		\citenamefont {{Bovo}}, \citenamefont {{Sheptyakov}}, \citenamefont
		{{Cervellino}}, \citenamefont {{Cava}}, \citenamefont {{Kenzelmann}},
		\citenamefont {{Spaldin}},\ and\ \citenamefont
		{{Fennell}}}]{2016PhRvB..93u4308R}%
	\BibitemOpen
	\bibfield  {author} {\bibinfo {author} {\bibfnamefont {M.}~\bibnamefont
			{{Ruminy}}}, \bibinfo {author} {\bibfnamefont {M.~N.}\ \bibnamefont
			{{Valdez}}}, \bibinfo {author} {\bibfnamefont {B.}~\bibnamefont
			{{Wehinger}}}, \bibinfo {author} {\bibfnamefont {A.}~\bibnamefont {{Bosak}}},
		\bibinfo {author} {\bibfnamefont {D.~T.}\ \bibnamefont {{Adroja}}}, \bibinfo
		{author} {\bibfnamefont {U.}~\bibnamefont {{Stuhr}}}, \bibinfo {author}
		{\bibfnamefont {K.}~\bibnamefont {{Iida}}}, \bibinfo {author} {\bibfnamefont
			{K.}~\bibnamefont {{Kamazawa}}}, \bibinfo {author} {\bibfnamefont
			{E.}~\bibnamefont {{Pomjakushina}}}, \bibinfo {author} {\bibfnamefont
			{D.}~\bibnamefont {{Prabakharan}}}, \bibinfo {author} {\bibfnamefont {M.~K.}\
			\bibnamefont {{Haas}}}, \bibinfo {author} {\bibfnamefont {L.}~\bibnamefont
			{{Bovo}}}, \bibinfo {author} {\bibfnamefont {D.}~\bibnamefont
			{{Sheptyakov}}}, \bibinfo {author} {\bibfnamefont {A.}~\bibnamefont
			{{Cervellino}}}, \bibinfo {author} {\bibfnamefont {R.~J.}\ \bibnamefont
			{{Cava}}}, \bibinfo {author} {\bibfnamefont {M.}~\bibnamefont
			{{Kenzelmann}}}, \bibinfo {author} {\bibfnamefont {N.~A.}\ \bibnamefont
			{{Spaldin}}},\ and\ \bibinfo {author} {\bibfnamefont {T.}~\bibnamefont
			{{Fennell}}},\ }\bibfield  {title} {\bibinfo {title} {{First-principles
				calculation and experimental investigation of lattice dynamics in the
				rare-earth pyrochlores R$_{2}$Ti$_{2}$O$_{7}$ (R =Tb,Dy,Ho)}},\ }\href
	{https://doi.org/10.1103/PhysRevB.93.214308} {\bibfield  {journal} {\bibinfo
			{journal} {\prb}\ }\textbf {\bibinfo {volume} {93}},\ \bibinfo {eid} {214308}
		(\bibinfo {year} {2016})}\BibitemShut {NoStop}%
	\bibitem [{\citenamefont {{Lan}}\ \emph {et~al.}(2015)\citenamefont {{Lan}},
		\citenamefont {{Ouyang}},\ and\ \citenamefont
		{{Song}}}]{2015AcMat..91..304L}%
	\BibitemOpen
	\bibfield  {author} {\bibinfo {author} {\bibfnamefont {G.}~\bibnamefont
			{{Lan}}}, \bibinfo {author} {\bibfnamefont {B.}~\bibnamefont {{Ouyang}}},\
		and\ \bibinfo {author} {\bibfnamefont {J.}~\bibnamefont {{Song}}},\
	}\bibfield  {title} {\bibinfo {title} {{The role of low-lying optical phonons
				in lattice thermal conductance of rare-earth pyrochlores: A first-principle
				study}},\ }\href {https://doi.org/10.1016/j.actamat.2015.03.004} {\bibfield
		{journal} {\bibinfo  {journal} {Acta Materialia}\ }\textbf {\bibinfo {volume}
			{91}},\ \bibinfo {pages} {304} (\bibinfo {year} {2015})}\BibitemShut
	{NoStop}%
	\bibitem [{\citenamefont {{Rodriguez}}\ \emph {et~al.}(2013)\citenamefont
		{{Rodriguez}}, \citenamefont {{Yaouanc}}, \citenamefont {{Barbara}},
		\citenamefont {{Pomjakushina}}, \citenamefont {{Qu{\'e}merais}},\ and\
		\citenamefont {{Salman}}}]{2013PhRvB..87r4427R}%
	\BibitemOpen
	\bibfield  {author} {\bibinfo {author} {\bibfnamefont {J.~A.}\ \bibnamefont
			{{Rodriguez}}}, \bibinfo {author} {\bibfnamefont {A.}~\bibnamefont
			{{Yaouanc}}}, \bibinfo {author} {\bibfnamefont {B.}~\bibnamefont
			{{Barbara}}}, \bibinfo {author} {\bibfnamefont {E.}~\bibnamefont
			{{Pomjakushina}}}, \bibinfo {author} {\bibfnamefont {P.}~\bibnamefont
			{{Qu{\'e}merais}}},\ and\ \bibinfo {author} {\bibfnamefont {Z.}~\bibnamefont
			{{Salman}}},\ }\bibfield  {title} {\bibinfo {title} {{Muon diffusion and
				electronic magnetism in Y$_{2}$Ti$_{2}$O$_{7}$}},\ }\href
	{https://doi.org/10.1103/PhysRevB.87.184427} {\bibfield  {journal} {\bibinfo
			{journal} {\prb}\ }\textbf {\bibinfo {volume} {87}},\ \bibinfo {eid} {184427}
		(\bibinfo {year} {2013})}\BibitemShut {NoStop}%
	\bibitem [{\citenamefont {Martelli}(2024)}]{martelli2024phonons}%
	\BibitemOpen
	\bibfield  {author} {\bibinfo {author} {\bibfnamefont {V.}~\bibnamefont
			{Martelli}},\ }\bibfield  {title} {\bibinfo {title} {Phonons bend to magnetic
			fields},\ }\href {https://doi.org/10.1038/s41567-023-02288-w} {\bibfield
		{journal} {\bibinfo  {journal} {Nature Physics}\ ,\ \bibinfo {pages} {1}}
		(\bibinfo {year} {2024})}\BibitemShut {NoStop}%
	\bibitem [{\citenamefont {Sharma}\ \emph {et~al.}(2024)\citenamefont {Sharma},
		\citenamefont {Valldor},\ and\ \citenamefont
		{Lorenz}}]{sharma_2024_12755212}%
	\BibitemOpen
	\bibfield  {author} {\bibinfo {author} {\bibfnamefont {R.}~\bibnamefont
			{Sharma}}, \bibinfo {author} {\bibfnamefont {M.}~\bibnamefont {Valldor}},\
		and\ \bibinfo {author} {\bibfnamefont {T.}~\bibnamefont {Lorenz}},\
	}\bibfield  {title} {\bibinfo {title} {{Data for ''Phonon thermal Hall effect in
				non-magnetic {Y${\rm _2}$Ti${\rm _2}$O${\rm _7}$}''}},\ }\href
	{https://doi.org/10.5281/zenodo.12755212} {10.5281/zenodo.12755212} (\bibinfo
	{year} {2024})\BibitemShut {NoStop}%
\end{thebibliography}

%
\end{document}